\documentclass{aa} 
\usepackage{graphics} 
\usepackage{graphicx} 
\usepackage[]{pict2e}
\usepackage{color, colortbl}
\usepackage{tikz}
\usepackage{pgfplots}
\usepackage{mathtools,amssymb}
\pgfplotsset{compat=1.18}

\usepackage{txfonts}

\begin{document}

\def\ammo{\rm NH_3}
\def\dammo{\rm NH_2D}
\def\ddammo{\rm NHD_2}
\def\ndthree{\rm ND_3}
\def\odammo{\rm oNH_2D}
\def\pdammo{\rm pNH_2D}
\def\oddammo{\rm oNHD_2}
\def\pddammo{\rm pNHD_2}
\def\diaz{\rm N_2H^+}
\def\ddiaz{\rm N_2D^+}
\def\htwo{\rm H_2}
\def\htwodplus{\rm H_2D^+}
\def\dtwohplus{\rm D_2H^+}
\def\dthreeplus{\rm D_3^+}
\def\hthreeplus{\rm H_3^+}
\def\percc{\rm cm^{-3}}
\def\dammoline{\rm NH_2D(1_{01}-0_{00})}
\def\ddammoline{\rm NHD_2(1_{11}-0_{00})}
\def\exgroundline{1_{11}-0_{00}}
\def\groundline{1_{01}-0_{00}}
\def\exline{1_{11}-1_{01}}
\def\kms{\rm km\,s^{-1}}
\def\ms{\rm m\,s^{-1}}
\def\persqcm{\rm cm^{-2}}
\def\pers{\rm s^{-1}}

\thicklines
\newcommand*\circled[1]{\tikz[baseline=(char.base)]{
            \node[shape=circle,draw,inner sep=2pt] (char) {#1};}}

\title{Statistical nuclear spin ratios of deuterated ammonia in the pre-stellar core L1544
  \thanks{
  The data were collected with the Atacama Pathfinder EXperiment (APEX; programme ID M-0112.F-9506A-2023), and with the 30 metre telescope of the Institut de radioastronomie millim{\'e}trique (IRAM). The APEX Project is led by the Max Planck Institute for Radio Astronomy at the ESO La Silla Paranal Observatory.}}
  
   \author{J. Harju
          \inst{1,2}
          \and
                    P. Caselli \inst{1}
          \and
                    O. Sipil{\"a} \inst{1}
          \and
                    S. Spezzano \inst{1}
           \and 
                    A. Belloche \inst{3}
             \and
          L. Bizzocchi \inst{4}     
          \and
          J. E. Pineda \inst{1}
                    \and
          E. Redaelli\inst{5,1}
         \and
          F. Wyrowski \inst{3}
    }

 \institute{Max-Planck-Institut f{\"u}r extraterrestrische Physik, Gie{\ss}enbachstra{\ss}e 1, D-85748 Garching, Germany 
   \and
   Department of Physics, P.O. Box 64, FI-00014, University of Helsinki, Finland
   \and
    Max-Planck-Institut f{\"u}r Radioastronomie,
    Auf dem H{\"u}gel 69, D-53121 Bonn, Germany
    \and
 Dipartimento di Chimica “Giacomo Ciamician”, Università di Bologna, via F. Selmi 2, I-40126 Bologna, Italy 
 \and
 European Southern Observatory, Karl-Schwarzschild-Stra{\ss}e 2, D-85748 Garching, Germany
    }

   \date{Received 29 April 2025; accepted 1 July 2025}

 \abstract
   {The relative abundances of the nuclear spin modifications of molecules contain information on their formation mechanism.}
   {We determined the ortho/para (o/p) ratios of $\dammo$ and $\ddammo$ in the archetypical pre-stellar core L1544.}
   {L1544 was observed in the two lowest rotational lines of ortho- and para-$\dammo$ using the Atacama Pathfinder EXperiment (APEX) and the IRAM 30\,m telescopes. The ground-state lines of ortho- and para-$\ddammo$ were observed with APEX. The distributions of chemical abundances in the core were predicted using a gas-grain chemistry model with two different scenarios concerning proton transfer reactions in the gas. One of the scenarios, the so-called full scrambling (FS), allows protons and deuterons to be completely mixed in the intermediate reaction complex before dissociation, whereas the other describes these reactions as proton or deuteron hops (PH). We also tested assumed abundance profiles independent of the chemistry models. Radiative transfer calculations were used to simulate the observed $\dammo$ and $\ddammo$ lines from the predicted and assumed abundance profiles. }
   {Our modelling efforts suggest that the ground-state lines of $\dammo$ and $\ddammo$ at the wavelength $\lambda=0.9$\,mm  that are observable with the same beam and in the same spectrometer band are the most reliable probes of the o/p ratios. Simulations using the PH reaction scheme show systematically better agreement with the observations than simulations with the FS model. Simulations using a broken power law abundance profile as a function of the gas density, which seems to agree with previous observations and models, give spin ratios that are close to the predictions of the PH scenario: o/p-$\dammo=2.85\pm0.05$, o/p-$\ddammo=2.10\pm0.06$ ($1\,\sigma$).}
   {The o/p ratios predicted by the PH scenario in the gas phase correspond to the nuclear spin statistical weights, that is, o/p-$\dammo=3$, o/p-$\ddammo=2$. In view of the fact that H and D atom addition reactions on grain surfaces also result in these ratios, it is reasonable to assume that the spin ratios of interstellar ammonia and its deuterated forms are in general equal to their statistical values.}

\maketitle

\section{Introduction}
\label{introduction}

Ammonia and its deuterated isotopologues are frequently used as probes of physical and chemical conditions in dense interstellar clouds through their rotation-inversion spectra at centimetre and millimetre wavelengths. These molecules contain two or three identical protons (fermions) or deuterons (bosons), and the symmetrisation postulate of quantum mechanics imposes certain restrictions on their complete internal wave functions, which describe the quantum state of an isolated molecule, including the electronic and rotational-vibrational states as well as the electronic and nuclear spins. For example, the complete internal wave function of $\dammo$ must change sign under the permutation of the two protons, while that of $\ddammo$ must remain unchanged by the interchange of the two deuterons. $\dammo$ and $\ddammo$ have both symmetric and anti-symmetric nuclear spin wave functions, and to comply with the symmetrisation postulate, these can only be associated with certain rotation-inversion states. The restrictions described above give rise to different nuclear spin modifications of $\ammo$, $\dammo$, $\ddammo$, and $\ndthree$, which can be treated as different chemical species. It is customary to use the Greek appellation `ortho' for the modification with the highest nuclear spin statistical weight (number of symmetrised nuclear spin functions that have appropriate symmetry) and `para' for the modification with the lowest weight. $\ndthree$ has a third symmetry species called `meta' (e.g. \citealt{bunker&jensen2005}; \citealt{2009JChPh.130p4302H}; \citealt{2016MNRAS.457.1535D}). 

Chemical models that distinguish between different spin species indicate that their relative abundances do not necessarily correspond to the nuclear spin statistical weights. Deviations from the statistical weights originate in chemical reactions that form ammonia if they proceed through an intermediate reaction complex where protons and deuterons can be completely mixed. The possible spin states of the reaction complex and the dissociation products are determined by selection rules that derive from the conservation of spin angular momentum and permutation symmetry (\citealt{2009JChPh.130p4302H}; \citealt{2015A&A...581A.122S}; \citealt{2016JChPh.145g4301S}; \citealt{2018MNRAS.477.4454H}). A practical reason to have a priori knowledge of the spin ratios is that it is not always easy to determine the abundances of all the spin modifications of a molecule and thus its total abundance directly from observations. For example, the fundamental transitions of ortho-$\ammo$ and ortho-$\ndthree$ are at unfavourable frequencies (572 and 615 GHz, respectively) for ground-based observations. The former line has been observed in the pre-stellar core L1544 with {\sl Herschel} but is difficult to interpret due to self-absorption \citep{2017A&A...603L...1C}. {Deviations from the statistical ortho/para (o/p) ratios of water and ammonia (the latter derived from NH$_2$) have been reported in comets (\citealt{2011ARA&A..49..471M}; \citealt{2016MNRAS.462S.124S}). These were previously thought to reflect the temperatures at which the ices in the comet nucleus were formed or condensed. Currently, the idea that possible deviations from statistical ratios are caused by gas-phase reactions in comae is perhaps more favoured (\citealt{2013ApJ...770L...2F}; \citealt{2018ApJ...857L..13H}; \citealt{2019MNRAS.487.3392F}). Experimental and theoretical studies suggest that water desorbing from cometary nuclei should have a statistical o/p ratio of 3 (\citealt{2018ApJ...857L..13H}; \citealt{2022A&A...663A..43C} and references therein). A couple of recent observational estimates for cometary comae find no significant deviation from this value when optical thickness effects are carefully taken into account (\citealt{2022A&A...663A..43C}; \citealt{2025PSJ.....6..139W}).} 

We have previously determined the o/p ratios of $\dammo$ and $\ddammo$ in two pre-stellar cores in Ophiuchus, H-MM1 and Oph D, using the Large APEX sub-Millimeter Array (LAsMA) multi-beam receiver of the Atacama Pathfinder EXperiment (APEX) telescope \citep{2024A&A...682A...8H}. These observations show, with a higher accuracy than previously achieved, that the mentioned spin ratios are within $20\%$ (taking $3\,\sigma$ limits into account) of their high-temperature statistical values, 3 and 2, respectively. The results were explained by a chemistry model that assumes the proton hop (PH) mechanism (see e.g. \citealt{2004JMoSp.228..635O}) in proton transfer and hydrogen abstraction reactions that contribute to the formation of ammonia \citep{2019A&A...631A..63S}. The alternative to PH is the so-called full scrambling (FS) mechanism, in which chemical reactions proceed via the formation of a relatively long-lived intermediate complex, where the light H and D nuclei can be mixed. According to our tests with both the PH and FS scenarios, the efficiency of ammonia desorption from grains does not affect its spin ratios in the gas. The insensitivity to the desorption rate was explained by a rapid reprocessing of released material in ion-molecule reactions, and this was considered to rule out the possibility that statistical spin ratios are caused by formation on grains. In contrast to PH, the FS model predicts clear deviations from the statistical spin ratios. In FS simulations, the o/p-$\dammo$ ratio generally decreases well below 3 or even below 2 in the gas, and o/p-$\ddammo$ always stays clearly above 2 (\citealt{2015A&A...581A.122S}; \citealt{2018MNRAS.477.4454H}).

The Ophiuchus molecular cloud complex, where H-MM1 and Oph\,D are located, is exposed to strong radiation from the Sco OB2 association and local B stars (e.g. \citealt{2020A&A...638A..74L}; \citealt{2022A&A...667A.163M}), and it is possible that the formation history and external conditions of the two cores have affected their composition in a way that current chemistry models cannot account for. Therefore, it seems helpful to repeat the measurement of the spin ratios in a quiescent environment.  To this end, we present observations of the ortho and para modifications of $\dammo$ and $\ddammo$ towards the pre-stellar core L1544, which lies on the outskirts of the Taurus  molecular cloud complex, far from strong sources of radiation and stellar winds.  The core is characterised by a high central density ($n(\htwo)>10^6\,\percc$) and a low central temperature ($\sim7$\,K; e.g., \citealt{2007A&A...470..221C}; \citealt{2019ApJ...874...89C}; \citealt{2022ApJ...929...13C}). An additional motivation for observing L1544 is that the column densities of ortho- and para-$\dammo$ ($\odammo$ and $\pdammo$) derived in the spectroscopic survey of \cite{2021JMoSp.37711431M} indicate an o/p-$\dammo$ ratio of $2.31\pm0.11$, which would agree with the predictions of FS models. This o/p ratio was derived from the ortho- and para-lines at 85.9\,GHz and 110.2\,GHz ($\lambda \sim 3.5$ and $2.7$\,mm) and includes uncertainty caused by the different beam sizes and the different Einstein coefficients and critical densities of the two transitions.  To confirm the low o/p-$\dammo$ ratio, and to also determine the o/p-$\ddammo$ ratio in this object, we observed the ground-state lines of both species at $\lambda\sim0.9$\,mm with APEX/LAsMA, and combined these with lines of the same molecules around $\lambda=3$\,mm, obtained with the 30\,m telescope of the Institut de radioastronomie millim{\'e}trique (IRAM). 

\section{Observations}
\label{observations}

\begin{figure}
\unitlength=1mm
\begin{picture}(80,80)(0,0)
\put(-15,0){
\begin{picture}(0,0) 
\includegraphics[width=11cm,angle=0]{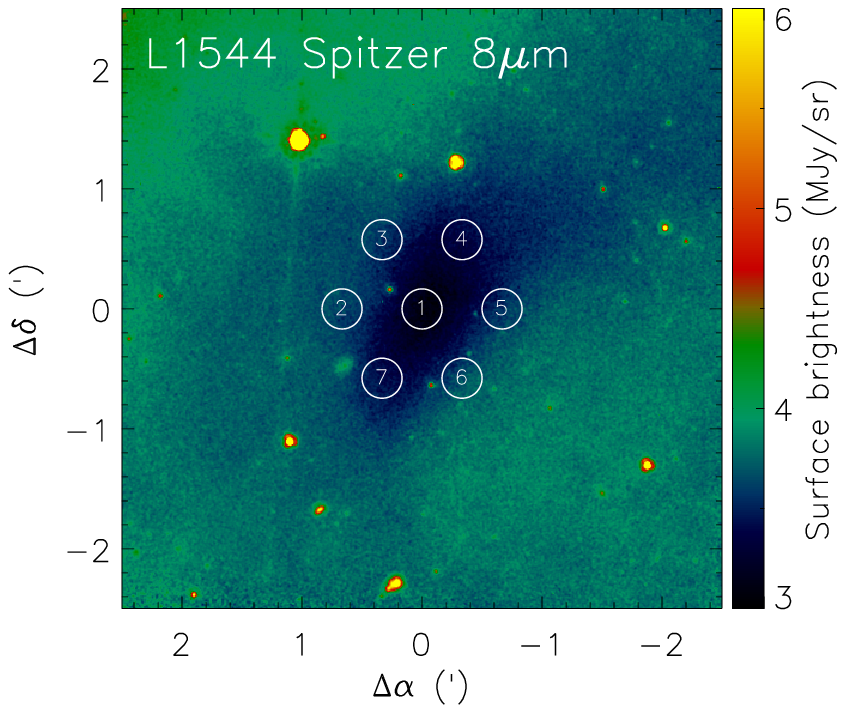}
\end{picture}}
\end{picture}  
   \caption{Footprint of the APEX/LAsMA array  on the $8\,\mu$m surface brightness images of L1544 observed by the {\sl Spitzer} satellite. The position of the central pixel is RA $05^{\rm h}04^{\rm m}17\fs1$, Dec. $+25\degr10\arcmin43\arcsec$ (J2000.0). The circle diameter corresponds to the HPBW of APEX at 334\,GHz. The core appears as a shadow against the weak mid-infrared background. The bright spots are foreground stars.}
   \label{lasma_footprint}
    \end{figure}
    
\subsection{APEX}
\label{apex}

The ground-state rotational lines of $\odammo$, $\pdammo$, ortho-$\ddammo$ ($\oddammo$), and para-$\ddammo$ ($\pddammo$) near 333 and 335\,GHz were observed towards L1544 with the LAsMA array \citep{2008SPIE.7020E..10G} on the APEX telescope \citep{2006A&A...454L..13G}. The observations were made in August 2023, under project number M-0112.F-9506A-2023.  These were single-point position switching observations with the central pixel of LAsMA pointed at the column density peak of L1544 (RA $05^{\rm h}04^{\rm m}17\fs1$, Dec. $+25\degr10\arcmin43\arcsec$ (J2000.0)). The reference position (OFF) was selected $10\arcmin$ to the equatorial east of the ON position. The array has 7 pixels separated by approximately $40\arcsec$. Figure~\ref{lasma_footprint} shows the footprint of the instrument superposed on the $8\,\mu$m map of the core measured by the InfraRed Array Camera (IRAC) of the Spitzer Space Telescope. The beam size of APEX is $18\farcs7$ (full width at half maximum) at 333\,GHz.  The 12\,m APEX telescope is located at an altitude of 5100\,m at Llano de Chajnantor in Chile. 

The backend was composed of fast Fourier transform spectrometers (FFTSs; called the FFTS4G) that covered two 4\,GHz intermediate frequency bands resolved into 65 536 spectral channels of width 61.03\,kHz (corresponding to $\sim 55\,\ms$). Lines of ortho- and para-$\dammo$ and $\ddammo$ were observed in the lower sideband of the receiver, which was centred at 334.145\,GHz. The frequencies of the observed transitions are listed in Table~\ref{transitions}. The energies of the upper transition level and the critical densities, $n_{\rm crit}$, of the transitions are also listed. Spectroscopic data were taken from \cite{2021JMoSp.37711431M}. The (de)excitation rate coefficients for $\dammo$ and $\ddammo$ in collisions with $\htwo$ were adopted from \cite{2014MNRAS.444.2544D} and \cite{2016MNRAS.457.1535D}.

The critical densities in Table~\ref{transitions} were estimated at 10\,K, using the optically thin formula presented by Shirley (\citeyear{2015PASP..127..299S}, Eq.~(4)). This formula takes into account all collisionally induced transitions starting from the upper transition level of the emission line. In addition to the downward collisions to the lower transition level, it is primarily the collisions $1_{01}\rightarrow1_{11}$ and $1_{11}-0_{00}$ that affect the critical densities of the $1_{01}\rightarrow0_{00}$ and $1_{11}\rightarrow1_{01}$ emission lines of $\odammo$ and $\pdammo$. For $\oddammo$ and $\pddammo$, the collisions $1_{11}\rightarrow1_{01}$ compete with collisions $1_{11}\rightarrow0_{00}$, which correspond to the observed ground-state lines.

\begin{table}
\caption[]{Observed transitions. }
\begin{tabular}{llrll} \hline
\multicolumn{2}{c}{Transition} & Freq. (MHz)$^a$ & $E_{\rm u}$ (K) & $n_{\rm crit}$ ($\percc$)$^b$ \\ \hline
\noalign{\smallskip}
\multicolumn{5}{c}{APEX} \\ \hline
\noalign{\smallskip}
$\odammo$ & $1_{01}^{\rm a}-0_{00}^{\rm a}$ & 332781.803 & 16.6 & $7.1\times10^4$ \\
$\pdammo$ & $1_{01}^{\rm s}-0_{00}^{\rm s}$ & 332822.521 & 16.0 & $6.6\times10^4$ \\
$\oddammo$ & $1_{11}^{\rm s}-0_{00}^{\rm s}$ & 335513.715 & 16.1 & $7.4\times10^4$ \\ 
$\pddammo$ & $1_{11}^{\rm a}-0_{00}^{\rm a}$ & 335446.240 & 16.3 & $8.4\times10^4$ \\ \hline
\noalign{\smallskip}
\multicolumn{5}{c}{IRAM 30\,m} \\ \hline
\noalign{\smallskip}
$\odammo$ & $1_{11}^{\rm s}-1_{01}^{\rm a}$ & 85926.258 & 20.7 &  $6.1\times10^4$\\
$\pdammo$ & $1_{11}^{\rm a}-1_{01}^{\rm s}$ & 110153.550 & 21.3 & $1.3\times10^5$\\
\hline
\noalign{\smallskip}
\end{tabular}

$^a$ The frequencies are adopted from \cite{2021JMoSp.37711431M}.

$^b$ Computed at 10\,K.
\label{transitions}
\end{table}

\subsection{IRAM 30\,m}
\label{iram}

The L1544 core was mapped in the 85.9\,GHz and 110.2\,GHz ortho- and para-$\dammo$ lines with the IRAM 30\,m telescope in the on-the-fly mode. The mapping was performed in 2015. The scan directions alternated between the east-west equatorial and the north-south direction. The separation between rows and columns of the 85.9\,GHz map was in general $5\arcsec$; the half-power beam width (HPBW) is $28\farcs6$ at this frequency, so the separation is $\sim 0.17\times{\rm HPBW}$. The area covered was $120\arcsec\times120\arcsec$. In the 110.2\,GHz map, the rows and columns were separated by $8\arcsec$ ($\sim0.36\times{\rm HPBW}$; the HPBW is $22\farcs3$), and the map covered an area of $\sim 160\arcsec\times160\arcsec$. Both maps were centred on the column density peak of the core, RA $05^{\rm h}04^{\rm m}17\farcs2$, Dec. $+25\degr10\arcmin43\arcsec$ (J2000.0). 

For imaging, the maps were interpolated and averaged to regular grids using the cell sizes  $5\arcsec$ and $8\arcsec$ for the 85.9\,GHz and 110.2\,GHz maps, respectively. As gridding function, we used a Bessel function of the first kind, $J_1$, tapered with a Gaussian (\citealt{2007A&A...474..679M}; \citealt{2008PASJ...60..445S}). In practice, the observation representing a certain grid position was obtained by accumulating all observations with a radius of one HPBW from this position, weighted by the function $J_1(\pi r/a)/(\pi r/a) \times \exp(-(r/b)^2)$, where $r$ is the distance from the grid position in units of 1/3 of the beam width, and the constants are $a=1.55$, $b=2.52$. This gridding function causes a small broadening of the effective beam. The effect is approximately 10\% for the 110.2\,GHz map, but negligible for the 85.9\,GHz map, where the cell size is small compared to the beam (see \citealt{2008PASJ...60..445S}, their Fig.~6). In addition to the gridding function, the accumulated spectra were weighted by the inverse square of the system noise temperature $T_{\rm SYS}$ (the integration time was the same for all samples).

The observations were performed with the IRAM Eight Mixer Receiver (EMIR) operating at millimetre wavelengths. The two $\dammo$ transitions required different spectral setups of the E90 channel of the receiver; the line at 85.9\,GHz was  observed in the lower-outer subband, whereas the 110.2\,GHz line was placed in the upper-outer subband. The EMIR receiver was connected to FFTSs. The FFTS back-end was configured to give 7.28\,GHz of instantaneous bandwidth and a channel width of 50\,kHz. The FFTS broadband spectra also covered the $1_{10}-1_{01}$ lines of ortho- and para-$\ddammo$ near 111\,GHZ. These lines were only marginally detected and were not useful in determining the o/p ratio.  

\subsection{ALMA}

{We included in the analysis the para-$\dammo(1_{11}-1_{01})$ map of L1544 that was observed in 2016 with the Atacama Large (Sub)millimeter Array (ALMA) and presented in \cite{2022ApJ...929...13C}. The observations consisted of a three-point mosaic using the 12\,m array, which is the main array, and a single pointing with the 7\,m array, the Atacama Compact Array (ACA). The project ID was 2016.1.00240.S (PI Caselli). Here, we used an image cube that was produced by using natural weighting of the combined (12\, m array and ACA) visibilities, and the CLEAN algorithm in the deconvolution of the dirty image. The final image was obtained  by convolving the CLEAN component image with a Gaussian beam of $2\farcs3\times1\farcs6$, which corresponds to the main lobe of the synthesised beam.  The rms noise in the central parts of the image is 0.15\,K per 42\,$\ms$ channel. The reduction procedure is described in Sect.\,2.1 of \cite{2022ApJ...929...13C}. The extent of p$\dammo$ emission was found to be approximately $1\arcmin\times0\farcm5$ (see Fig.\,A.1 in \citealt{2022ApJ...929...13C}, and Fig.~\ref{pnh2d_alma} below). } 

\subsection{Observed spectra and maps}
\label{spectra_maps}

\begin{figure*}
\centering
\unitlength=1mm
\begin{picture}(160,65)(0,0)
\put(-5,0){
\begin{picture}(0,0) 
\includegraphics[width=8cm,angle=0]{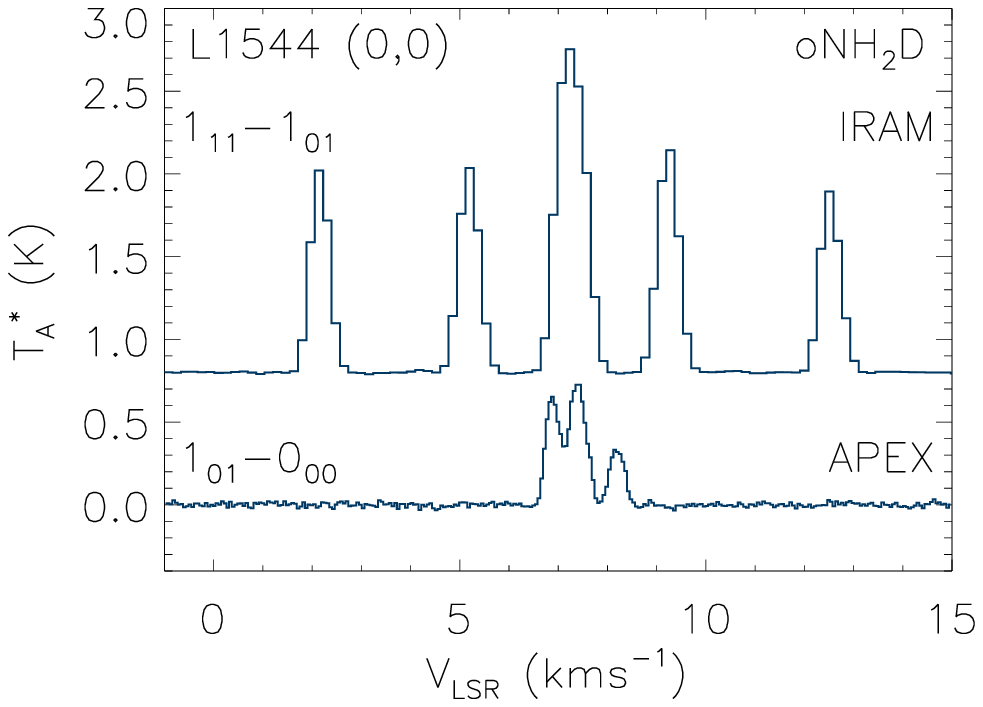}
\end{picture}}
\put(85,0){
\begin{picture}(0,0) 
\includegraphics[width=8cm,angle=0]{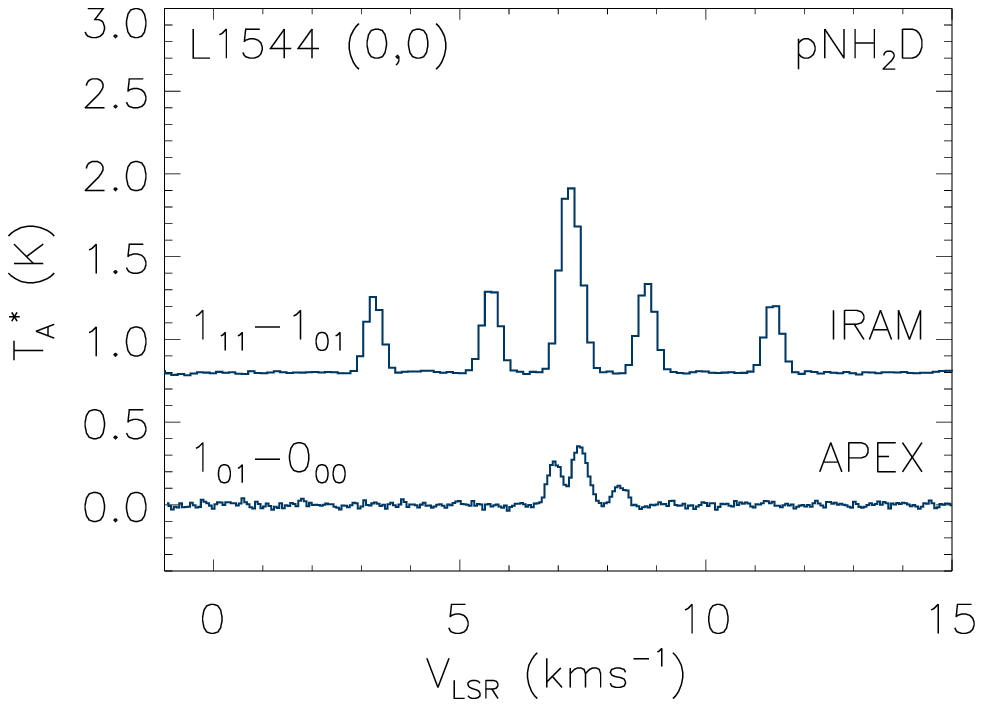}
\end{picture}}
\end{picture}  
\caption[]{$\dammo$ spectra observed with IRAM and APEX towards the column density peak of L1544, which is the position of pixel \circled{1} in Fig.~\ref{lasma_footprint}. The IRAM spectra are shifted up by 0.8\,K for clarity.}
\label{peak_nh2d_spectra}
\end{figure*}

\begin{figure}
\unitlength=1mm
\begin{picture}(90,65)(0,0)
\put(0,0){
\begin{picture}(0,0) 
\includegraphics[width=8cm,angle=0]{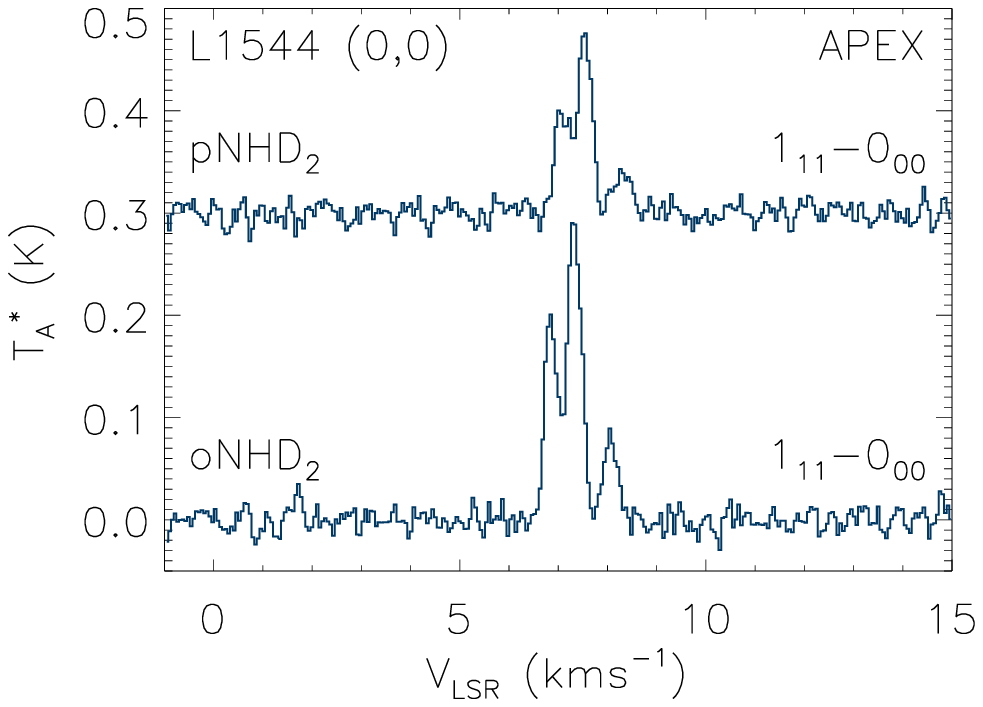}
\end{picture}}
\end{picture}
\caption[]{$\ddammo$ spectra observed with APEX towards the column density peak of L1544 ( pixel \circled{1} in Fig.~\ref{lasma_footprint}). The $\pddammo$ spectrum is shifted up by 0.3\,K for display purposes.}
\label{peak_nhd2_spectra}
\end{figure}

\begin{figure*}
\centering
\unitlength=1mm
\begin{picture}(160,70)(0,0)
\put(-5,70){
\begin{picture}(0,0) 
\includegraphics[width=7cm,angle=270]{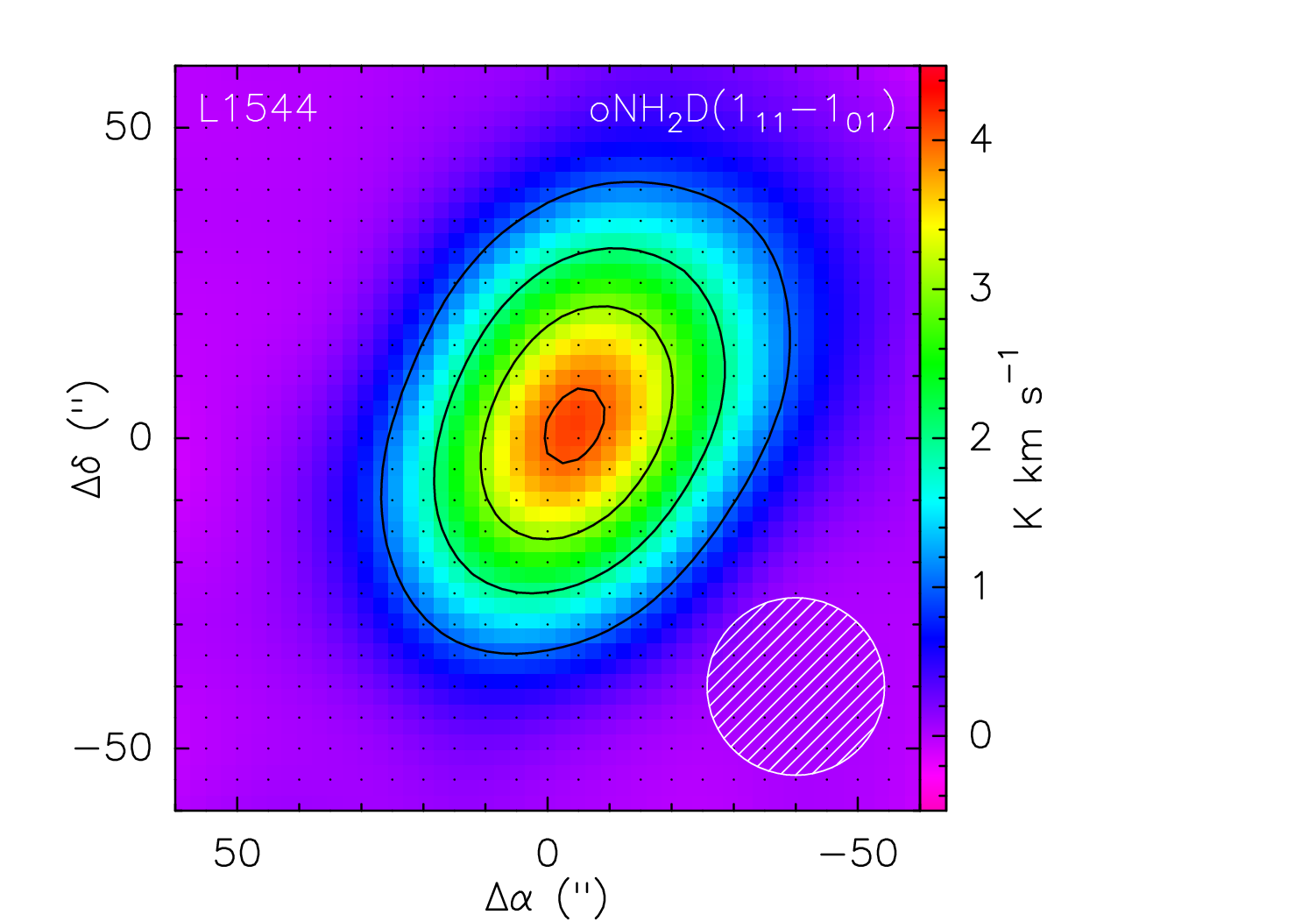}
\end{picture}}
\put(85,70){
\begin{picture}(0,0) 
\includegraphics[width=7cm,angle=270]{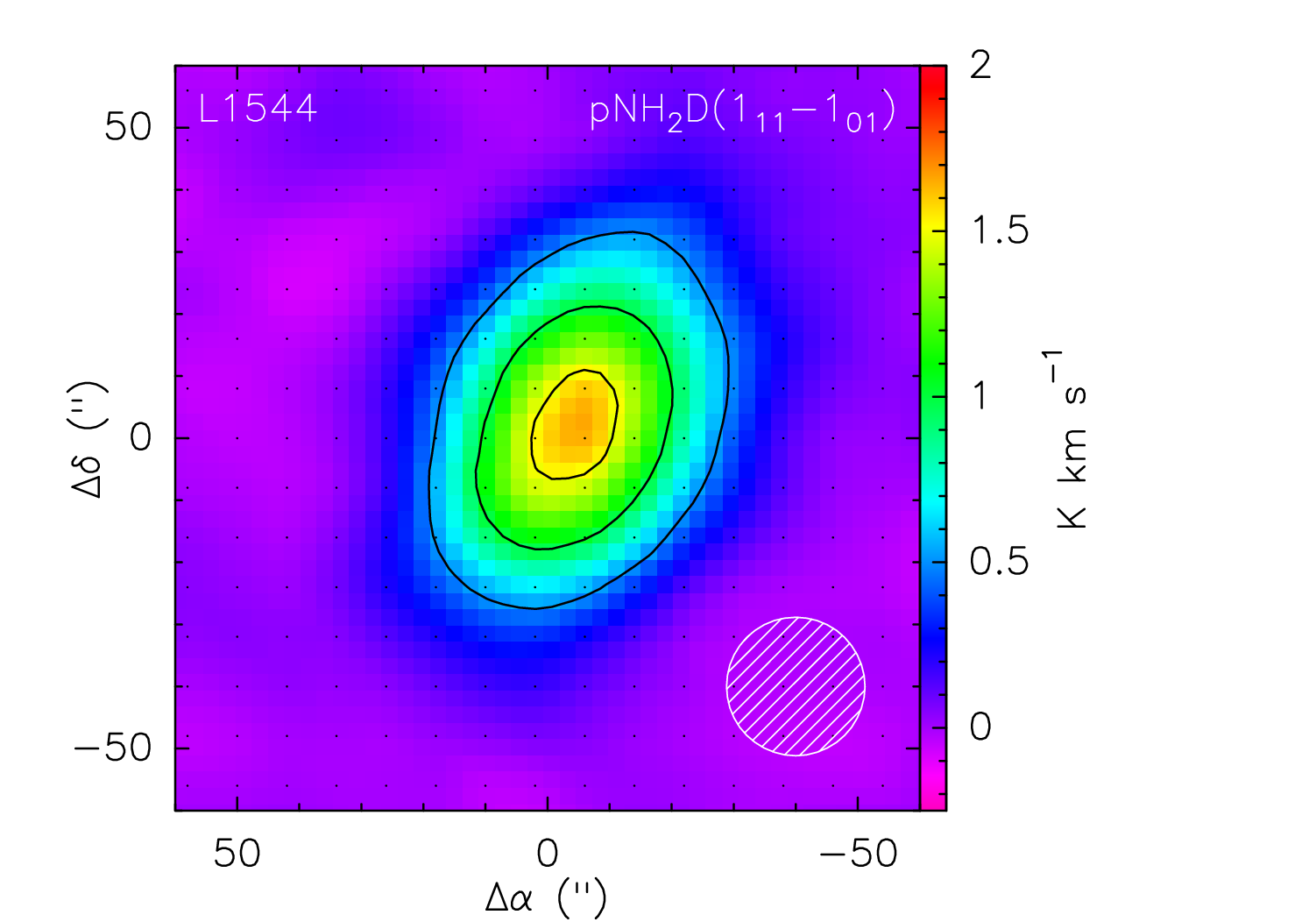}
\end{picture}}
\end{picture}  
\caption[]{Integrated $J_{K_a,K_b}=1_{11}-1_{01}$ line intensity maps of $\odammo$ and $\pdammo$ towards L1544 observed with the IRAM 30\,m telescope. The intensity is on the $T_{\rm A}^*$ scale. The beam sizes at the line frequencies (85.9\,GHz and 110.2\,GHz) are indicated in the bottom right.}
\label{nh2d_maps}
\end{figure*}

The $\dammo$ and $\ddammo$ spectra observed with APEX and IRAM towards the $\htwo$ column density maximum of L1544 are shown in Figs.~\ref{peak_nh2d_spectra} and \ref{peak_nhd2_spectra}. The APEX spectra correspond to pixel \circled{1} of the LAsMA array. The integrated intensity maps of the 85.9\,GHz and 110.2\,GHz lines of $\dammo$ observed with the 30\,m telescope are shown in Fig.~\ref{nh2d_maps}. Spectra and maps are presented on the antenna temperature, $T_{\rm A}^*$, scale. 

\begin{centering}
\begin{table*}
\centering
\caption[]{Line parameters and column densities from Gaussian fits to the hyperfine structure.}
\label{line_parameters}
  \begin{tabular}{ccccccc} \hline
   \noalign{\smallskip}
    \noalign{\smallskip}  
    \multicolumn{7}{c}{$\odammo$} \\
     \noalign{\smallskip}
transition & peak $T_{\rm MB}$ & $\upsilon_{\rm LSR}$ & $\Delta \upsilon$ & $T_{\rm ex}$ & $\tau$ & $N_{\rm tot}$ \\
$J_{K_a,K_c}$ & (K)     &  ($\kms$)     & ($\kms$) &   (K)  &    & ($10^{14}\, \persqcm$) \\ \hline
 \noalign{\smallskip}
$1_{11} - 1_{01}$ & 2.291 (0.005) & 7.263 (0.001) & 0.359 (0.001) & 5.31 (0.01) & 9.05 (0.04)  &4.13 (0.04) \\
$1_{01} - 0_{00}$ & 0.965 (0.016) & 7.319 (0.001) & 0.282 (0.003) & 5.69 (0.02) & 5.71 (0.20) & 3.55 (0.16) \\ \hline
 \noalign{\smallskip}    \noalign{\smallskip}
    \multicolumn{7}{c}{$\pdammo$} \\
     \noalign{\smallskip}
transition & $T_{\rm MB}$ & $\upsilon_{\rm LSR}$ & $\Delta \upsilon$ & $T_{\rm ex}$ & $\tau$ & $N_{\rm tot}$ \\
$J_{K_a,K_c}$ & (K)     &  ($\kms$)     & ($\kms$) &   (K)  &    & ($10^{14}\,\persqcm$) \\ \hline
 \noalign{\smallskip}
$1_{11} - 1_{01}$ & 1.341 (0.007) & 7.254 (0.001) & 0.326 (0.002) & 4.86 (0.02) & 3.67 (0.06) & 1.52 (0.07) \\
$1_{01} - 0_{00}$ & 0.470 (0.017) & 7.362 (0.003) & 0.293 (0.007) & 4.98 (0.11) & 2.53 (0.34) & 1.54 (0.33) \\ \hline
 \noalign{\smallskip}    \noalign{\smallskip}
    \multicolumn{7}{c}{$\oddammo$} \\
     \noalign{\smallskip}
transition & $T_{\rm MB}$ & $\upsilon_{\rm LSR}$ & $\Delta \upsilon$ & $T_{\rm ex}$ & $\tau$ & $N_{\rm tot}$ \\
$J_{K_a,K_c}$ & (K)     &  ($\kms$)     & ($\kms$) &   (K)  &    & ($10^{13}\,\persqcm$) \\ \hline
 \noalign{\smallskip}
$1_{11} - 0_{00}$ & 0.367 (0.010) & 7.254 (0.002) & 0.323 (0.006) & 5.70 (0.37) & 0.89 (0.22) & 2.91 (1.17) \\ \hline 
 \noalign{\smallskip}    \noalign{\smallskip}
    \multicolumn{7}{c}{$\pddammo$} \\
     \noalign{\smallskip}
transition & $T_{\rm MB}$ & $\upsilon_{\rm LSR}$ & $\Delta \upsilon$ & $T_{\rm ex}$ & $\tau$ & $N_{\rm tot}^*$ \\
$J_{K_a,K_c}$ & (K)     &  ($\kms$)     & ($\kms$) &   (K)  &    & ($10^{13}\,\persqcm$) \\ \hline
 \noalign{\smallskip}
$1_{11} - 0_{00}$ & 0.221 (0.009) & 7.466 (0.004) & 0.353 (0.010) & 7.46 (4.48) & - & 1.24 (0.14)$^*$ \\ \hline 
  \noalign{\smallskip}   
  \end{tabular}
  
  $^*$ Column density using the optically thin approximation assuming $T_{\rm ex}(\rm para)=T_{\rm ex}(\rm ortho)$. 
\end{table*}
\end{centering}

The line parameters and column densities derived from Gaussian fits to the hyperfine structure of the lines observed towards the column density peak of L1544 are presented in Table~\ref{line_parameters}. {  The observed spectra were converted to the main-beam brightness temperature scale, $T_{\rm MB}$, using the beam efficiencies provided by the telescope teams. In column density estimates, we assumed uniform beam filling, line-of-sight homogeneity,} and that the excitation temperature, $T_{\rm ex}$, is constant for all rotational transitions of the molecule. Furthermore, the calculation involved the assumption that populations of hyperfine states at a certain rotational level are proportional to their statistical weights. The strong quadrupole interaction of the N nucleus with the electric field of the electrons gives rise to five distinct components of the $J_{K_a,K_c}=1_{11}-1_{01}$ lines of $\dammo$, and three components of the ground-state lines ($1_{01}-0_{00}$ and $1_{11}-0_{00}$) of $\dammo$ and $\ddammo$.  These are further split into several unresolved components owing to quadrupole interaction of the deuteron, and weaker spin-rotation and spin-spin couplings (\citealt{2006A&A...449..855C}; \citealt{2021JMoSp.37711431M}). The line width listed in Table~\ref{line_parameters} refers to the intrinsic width of an individual hyperfine component. The derived local standard of rest velocities of the  $1_{11}-0_{00}$ lines of $\oddammo$ and $\pddammo$ differ by approximately $210\,\ms$, whereas velocity difference of the $\odammo$ and $\pdammo$ lines in the same band is only $43\,\ms$. This suggests that the rest frequencies of the $\ddammo$ lines are not as accurate as those of the $\dammo$ lines. The velocity difference $210\,\ms$ corresponds to $\sim 240$\,kHz.   

The strong rotation-inversion lines of $\dammo$ at $\lambda=3$\,mm with five hyperfine groups have very high signal-to-noise ratios, and the formal errors of the column densities of $\dammo$ derived from these lines are clearly smaller than the estimates from the $\lambda=0.9$\,mm lines. Although the column densities of $\pdammo$ from the $1_{11}-1_{01}$ and $1_{01}-0_{00}$ lines agree, the total $\odammo$ column densities derived from the two ortho-lines differ by approximately 15\%.  This is due in part to the different beam sizes for the two o$\dammo$ lines ($29\arcsec$ vs $19\arcsec$). For the two p$\dammo$ lines the difference is smaller ($22\arcsec$ vs $19\arcsec$). The o/p-$\dammo$ ratio from the IRAM observations is $2.7\pm0.2$, whereas from the APEX spectra we obtain o/p-$\dammo=2.3\pm 0.5$.  The hyperfine structure fit to the p$\ddammo(1_{11}-0_{00})$ line does not give reliable estimates for the $T_{\rm ex}$ and $\tau$,  and we estimated the column density using the optically thin approximation, adopting the $T_{\rm ex}$ derived for the ortho-line. The resulting  o/p-$\ddammo$ ratio is $2.3\pm1.2$. {  Depending on the distribution of the gas density and molecular abundances, the effective source size can be different for each observed line. This causes an uncertainty in the beam-source coupling efficiency, which propagates to $T_{\rm ex}$.} In addition to the uncertainties involved in the determination of $T_{\rm ex}$ and $\tau$, especially when the line only has three resolved components, the assumption of a constant $T_{\rm ex}$ along the line of sight in the presence of large density gradients is an additional source of uncertainty. In what follows, we estimate the column densities and the fractional abundances of the observed molecules applying radiative transfer calculations to a spheroidal model of the L1544 core. 

\section{Core model}
\label{core_model}

We adopted the density and gas and dust temperature distributions from the model used for L1544 by \cite{2019MNRAS.487.1269S} and \cite{2023A&A...680A..87R}, which, in turn, was developed from a model derived by \citet[see Fig.~2 of \citealt{2023A&A...680A..87R}]{2010MNRAS.402.1625K}. However, because the core is clearly elongated, we deformed the originally spherical model to a prolate spheroid.  A Gaussian fit to the Spitzer $8\,\mu$m image of the core gave an axis ratio of 2.0 and a long axis tilt angle of $-36\fdg1$ from the declination axis. For the purpose of radiative transfer calculations, we constructed a cubical grid with an edge length of $\sim 180\arcsec$ or 30,600 au. The transformation to the spheroidal model was done so that the volume within an isosurface (a surface representing constant values of the density, temperature, etc.) is the same as that of the corresponding isosurface of the spherical model. Consequently, the gas properties at point $(x,y,z)$ in the grid, measured from the origin of the principal axis system, correspond to those of the spherical model at radius $r$, which is obtained from $r=2^{1/3}\{x^2 + y^2 + z^2/\gamma^2\}^{1/2}$, where $\gamma \equiv c/a$ is the axis ratio of the spheroid (here $\gamma=2$). 

We first tested how well the observed spectra can be reproduced by assuming that the fractional abundances of $\dammo$ and $\ddammo$ are constant along the line of sight. Line simulations were performed using the Line radiative transfer with the OpenCL (LOC) program \citep{2020A&A...644A.151J}. {  The hyperfine structure of the lines was modelled by assuming local thermodynamic equilibrium between the components, that is, the effects of line overlap were not taken into account.} The fractional abundances were determined by comparing the simulated spectra with the observed ones; the method that applies minimum chi-square estimation is described in Sect.\,3.2 of \cite{2024A&A...682A...8H}. {  Only the statistical uncertainties based on the noise of the observed spectra are taken into account in this method.} The best-fit abundances and their $1\,\sigma$ uncertainties obtained in the centre of the map (0,0) and the second strongest position  $(-23\arcsec,35\arcsec)$ are listed in Table~\ref{chi2_fracs}. In the latter position, the $\ddammo$ lines are weak, and only the $\dammo$ abundances are given. The $\dammo$ and $\ddammo$ spectra observed in the seven LAsMA positions, including (0,0) and $(-23\arcsec,35\arcsec)$, are shown in Figs.~\ref{lasma_dammo} and \ref{lasma_ddammo}. The derived fractional $\dammo$ abundance is clearly higher towards the offset position than towards the centre of the core. We note that the $N(\htwo)$ estimate used for the fractional abundance comes from the spheroidal core model, which may miss a local condensation near position $(-23\arcsec,35\arcsec)$. The other alternative is that the assumption of constant abundances throughout the core is not true, which would also render the assumption of line-of-sight constancy illogical. In what follows, we apply a chemistry model to predict the abundance distributions in the core and simulate spectra from the obtained models.    

\begin{centering}
\begin{table}
      \caption[]{Fractional abundances, $X$, and abundance ratios in L1544 derived from the LAsMA spectra in pixels \circled{1} (0,0) and \circled{4} ($-23\arcsec,35\arcsec$) assuming that $X$ is constant throughout the core.}
         \label{chi2_fracs}
         \begin{tabular}{l|c|c}
          \noalign{\smallskip}
          \multicolumn{1}{c}{} &
          \multicolumn{1}{c}{(0,0)} & 
          \multicolumn{1}{c}{($-23\arcsec,35\arcsec$)} \\ 
           & ($\times10^{-9}$)$^a$  &  ($\times10^{-9}$)$^a$ \\ \hline
           \noalign{\smallskip}
        $X(\odammo)$ & $5.07 (0.05)$ & $12.7 (0.2)$\\
        $X(\pdammo)$ &  $1.62 (0.03)$ & $4.30 (0.27)$ \\
        $X(\oddammo)$ & $1.04 (0.02)$ & \\
        $X(\pddammo$ &  $0.481 (0.009)$ & \\ \hline
        \noalign{\smallskip}
        $X(\dammo)$ &  $6.69 (0.06)$ & $17.0 (0.3)$\\ 
        $X(\ddammo)$ &  $1.52 (0.03)$ & \\ \hline \hline
        \noalign{\smallskip}
        $\dammo/\dammo$ &  $0.23 (0.01)$ & \\
        \noalign{\smallskip}
        o/p$\dammo$ &  $3.13 (0.07)$ & $2.95 (0.19)$ \\
        o/p$\ddammo$ & $2.15 (0.06)$  \\ \hline
         \end{tabular}

         $^a$ Factor of the fractional abundances, $X.$
       \end{table}
\end{centering}  

\section{Chemistry model}
\label{chemistry_model}

For chemical simulations, we used the gas-grain chemical model {\sl pyRate} described in \citet{2019A&A...631A..63S}, \citet{2023A&A...680A..87R}, and \citet{2024A&A...682A...8H}, employing the PH and FS scenarios for proton donation and abstraction reactions. In the PH network, proton transfer reactions involving O, C, and N are assumed to proceed via the exchange of one proton (or deuteron) between the reactants, while in the FS network multiple atom exchanges are possible. However, the $\hthreeplus + \htwo$ reaction system and its deuterated variants form an exception in the PH network. For the reactions that involve only the H or D nuclei, the rate coefficients and branching ratios adopted from \cite{2009JChPh.130p4302H}, follow the FS scenario in both networks. Some deuterated variants of the $\hthreeplus + \htwo$ system were recently studied by \cite{2024A&A...692A.121J} through experiments and simulations. The results indicate that the growth of the abundances of $\dtwohplus$, $\dtwohplus$, and $\dthreeplus$ is suppressed more than previously thought in warm conditions by hydrogenation reactions. Since we are dealing with very cold gas in L1544, we did not make any changes to the original reaction set of \cite{2009JChPh.130p4302H}. 

The efficiency of chemical desorption was assumed to be constant, with 1\% exothermic surface reactions leading to desorption. The dust grains were assumed to have uniform size characterised by the grain radius $a$ and the density of the material $\rho$, assumed to be $2.5\,{\rm g}\,\percc$. The simulations were performed using the grain radii $a=0.1$, 0.2, and $0.3\,\mu$m. The value $a=0.1\,\mu$m corresponds to a grain surface area of $5.0\times10^{-22}\,\persqcm$ per H atom, with the adopted material density and the dust-to-gas mass ratio (0.01; \citealt{2020A&A...640A..94S}). The grain surface area is inversely proportional to the grain radius for particles of uniform size. We also varied the cosmic-ray ionisation rate, testing the values 0.5,1,2,4, and 8 times $1.3\times10^{-17}\,\pers$. 

We assumed that the gas has been processed under low-density `dark cloud' conditions ($n(\htwo)=10^4\,\percc$, $T=10$\,K, $A_{\rm V}=5^{\rm mag}$) for half a million years before the `dense core' phase. The initial abundances used for the first simulation step are the same as those listed in Table~1 of \cite{2023A&A...680A..87R}. The chemical abundances attained at the end of the dark cloud phase were in turn used as initial values of the dense core phase.  The abundances in the dense core were calculated for concentric spherical shells using the one-dimensional model described in \cite{2023A&A...680A..87R}. These were interpolated to the spheroidal model for spectral line simulations using the transformation described above. 

\begin{figure*}
\centering
\unitlength=1mm
\begin{picture}(160,65)(0,0)
\put(-10,0){
\begin{picture}(0,0) 
\includegraphics[width=9cm,angle=0]{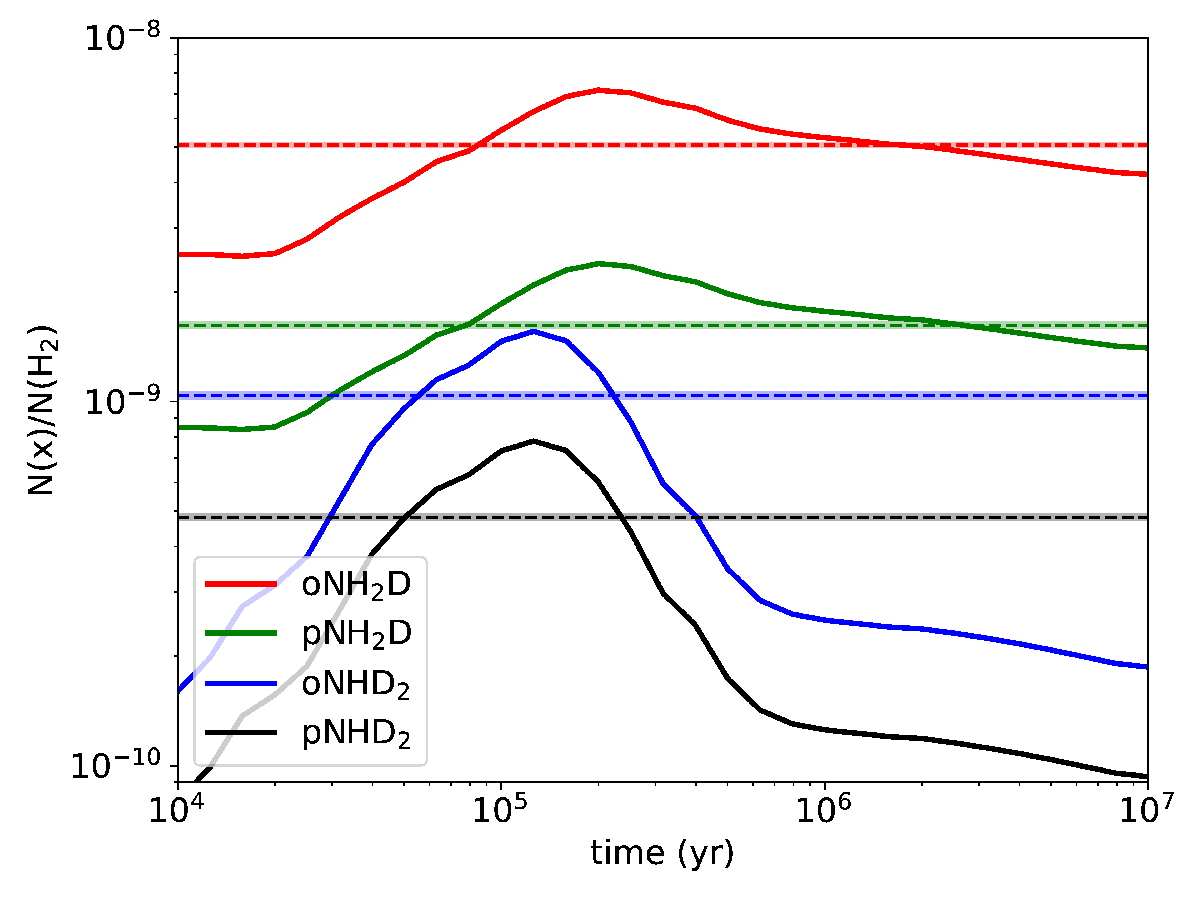}
\end{picture}}
\put(80,0){
\begin{picture}(0,0) 
\includegraphics[width=9cm,angle=0]{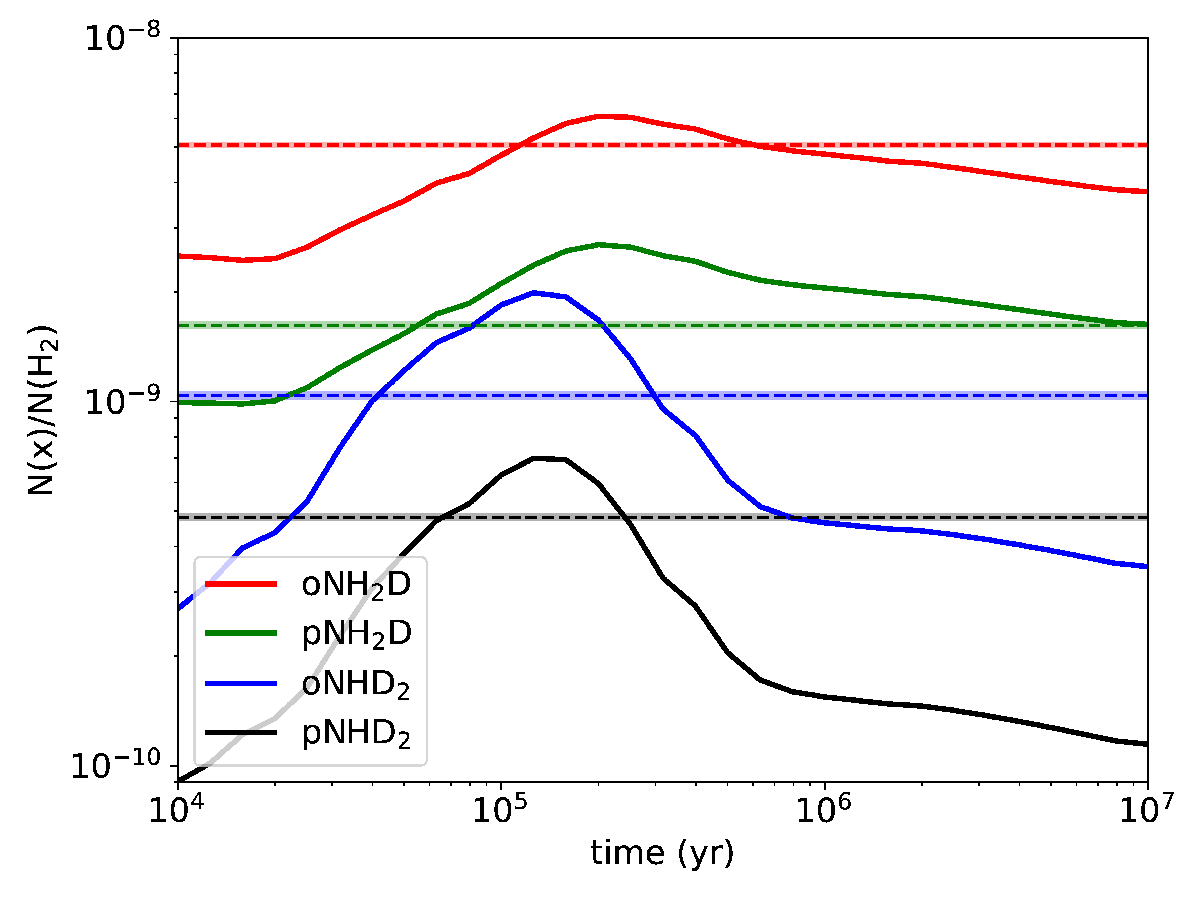}
\end{picture}}
\put(7,60){\large \bf (a)}
\put(97,60){\large \bf (b)}
\put(67,60){\large \bf }
\put(158,60){\large \bf }
\end{picture}  
\caption[]{Evolution of the fractional abundances of $\odammo$, $\pdammo$, $\oddammo$, and $\pddammo$ in the core model. The abundances correspond to the column density ratios $N({\rm mol})/N(\htwo)$ towards the centre of the core, smoothed to the resolution of the $18\farcs7$ APEX beam at 333 GHz. Predictions from the PH scenario in the chemical network  (a) and the FS scenario (b). In this simulation, $a=0.2\,\mu$m and $\zeta_{\htwo}=2.6\times10^{-17}\,\pers$. The horizontal dashed lines and thick semi-transparent lines represent the fractional abundances and their formal $1\sigma$ errors derived towards (0,0) in Sect.~\ref{core_model}. These abundances are listed in Table~\ref{chi2_fracs}.}
\label{x_vs_time}
\end{figure*}

\begin{figure*}
\centering
\unitlength=1mm
\begin{picture}(160,65)(0,0)
\put(-10,0){
\begin{picture}(0,0) 
\includegraphics[width=9cm,angle=0]{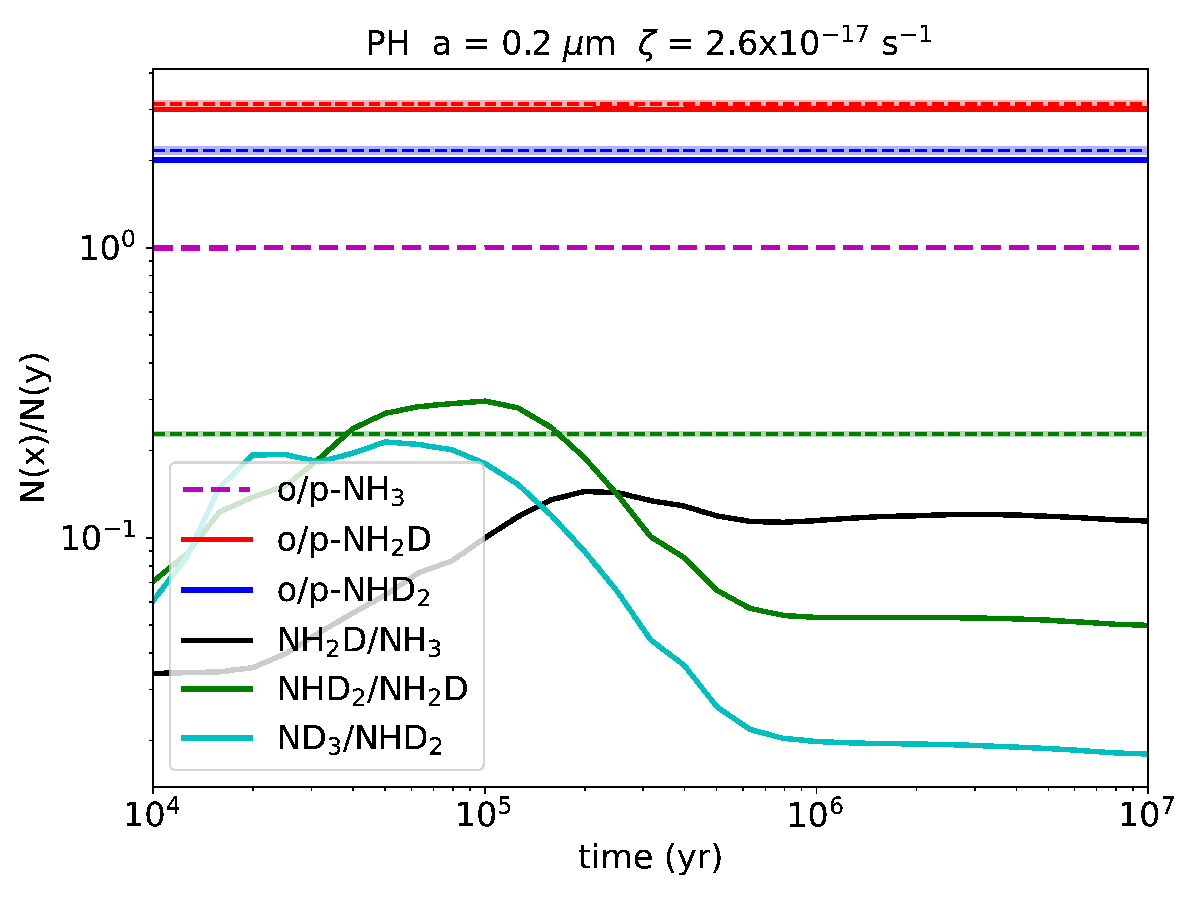}
\end{picture}}
\put(80,0){
\begin{picture}(0,0) 
\includegraphics[width=9cm,angle=0]{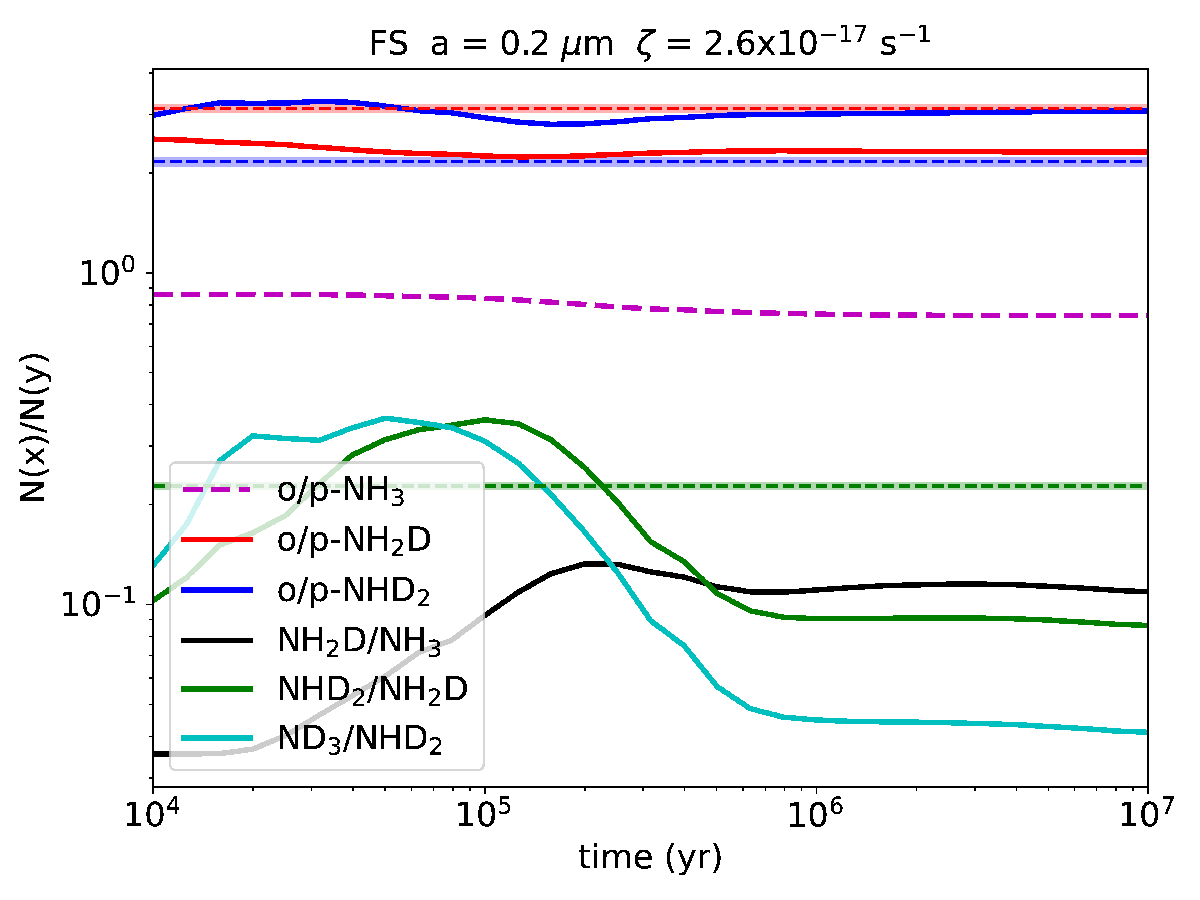}
\end{picture}}
\put(7,50){\large \bf (a)}
\put(97,50){\large \bf (b)}
\put(67,50){\large \bf }
\put(158,50){\large \bf }
\end{picture}  
\caption[]{{  Evolution of the o/p-$\ammo$, o/p-$\dammo$, and o/p-$\ddammo$ ratios} and the fractionation ratios $\dammo/\ammo$, $\ddammo/\dammo$, and $\ndthree/\ddammo$ according to the PH (a) and FS (b) models visualised in Fig.~\ref{x_vs_time}. The fractional abundances used for the ratios are calculated in the same way as those shown in Fig.~\ref{x_vs_time}. The horizontal straight dashed lines and semi-transparent thick lines represent the abundance ratios derived towards (0,0) from APEX observations (Sect.~\ref{core_model}). Observational data for $\ammo$ and $\ndthree$ are not available at the same resolution.}
\label{ops_dfracs}
\end{figure*}

The evolution of the average fractional abundances of $\dammo$ and $\ddammo$ in the FS and PH models is shown in Fig.~\ref{x_vs_time}, with the ortho- and para-species separated. Here, the grain radius is $a=0.2\,\mu$m, and the cosmic-ray ionisation rate is $\zeta_{\htwo}=2.6\times10^{-17}\,\pers$. The curves represent the column density ratios $N({\rm mol})/N(\htwo)$ towards the centre of the core. These column densities are derived from the spheroidal models and convolved with the telescope beam. The ratios are in this sense comparable to the estimates presented in Table~\ref{chi2_fracs} for the (0,0) position. In Fig.~\ref{ops_dfracs} we show the o/p ratios of $\ammo$,  $\dammo$, and $\ddammo$, and the fractionation ratios $\dammo/\ammo$, $\ddammo/\dammo$, and $\ndthree/\ddammo$ as functions of time predicted by the same PH and FS models as shown in Fig.~\ref{x_vs_time}. The PH model predicts constant o/p ratios corresponding to their high-temperature statistical values, whereas in the FS model the ratios show some time variation. One can see that the predicted values of the o/p-$\dammo$ and o/p-$\ddammo$ ratios are almost swapped between the PH and FS models. In agreement with the simulations of \cite{2013ApJ...770L...2F} and \cite{2014A&A...562A..83L}, the o/p ratio of $\ammo$ remains below 1 in the FS model. The deuterium fractionation ratios will be discussed briefly in Sect.~\ref{op_ratios}.    

\section{Simulated spectra}
\label{simulated_spectra}

\begin{figure*}
\centering
\unitlength=1mm
\begin{picture}(160,120)(0,0)
\put(-5,60){
\begin{picture}(0,0) 
\includegraphics[width=8cm,angle=0]{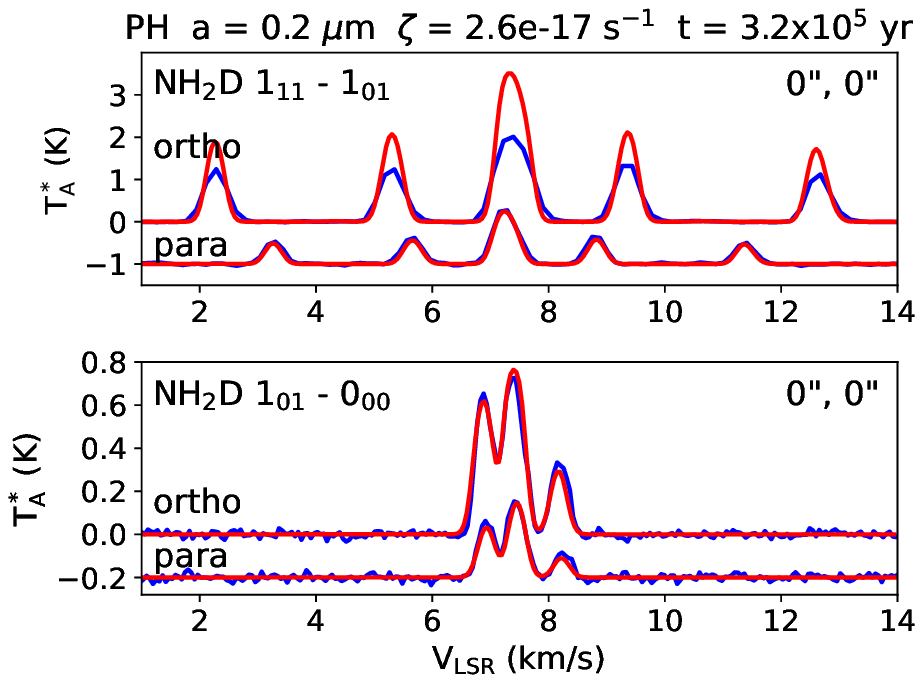}
\end{picture}}
\put(85,60){
\begin{picture}(0,0) 
\includegraphics[width=8cm,angle=0]{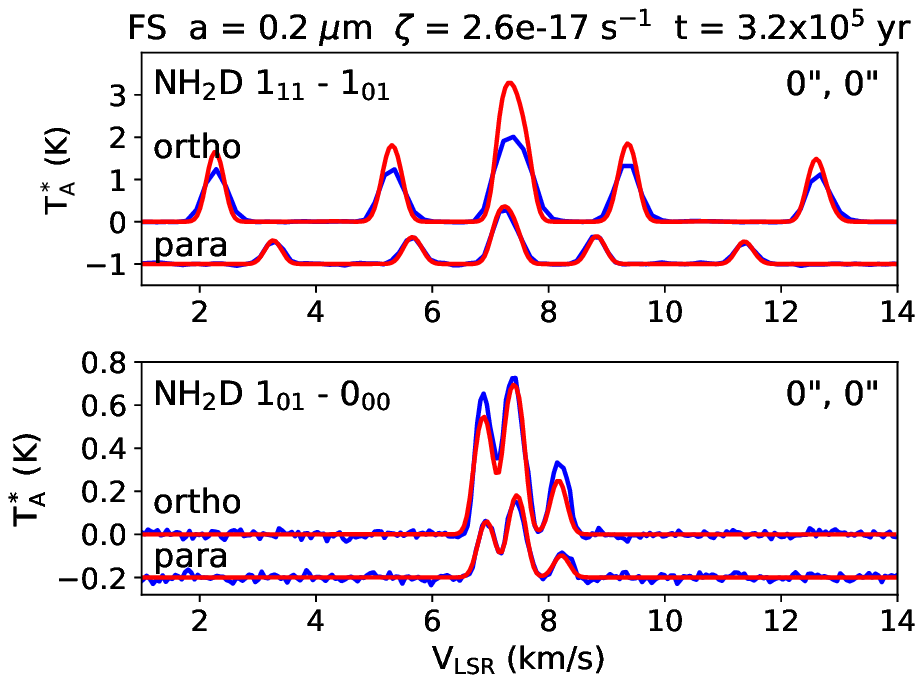}
\end{picture}}

\put(-5,0){
\begin{picture}(0,0) 
\includegraphics[width=8cm,angle=0]{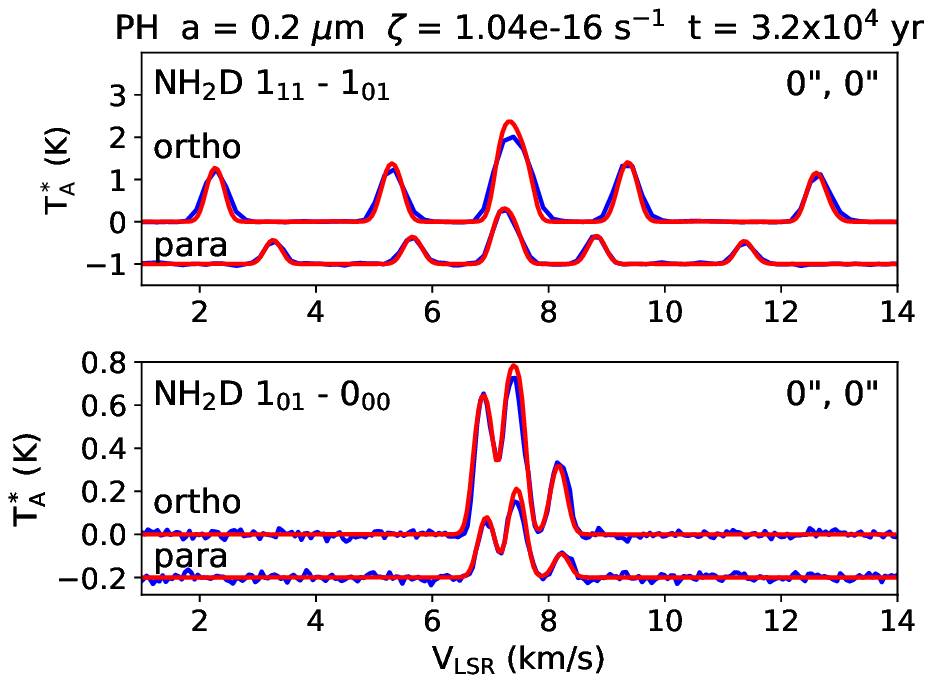}
\end{picture}}
\put(85,0){
\begin{picture}(0,0) 
\includegraphics[width=8cm,angle=0]{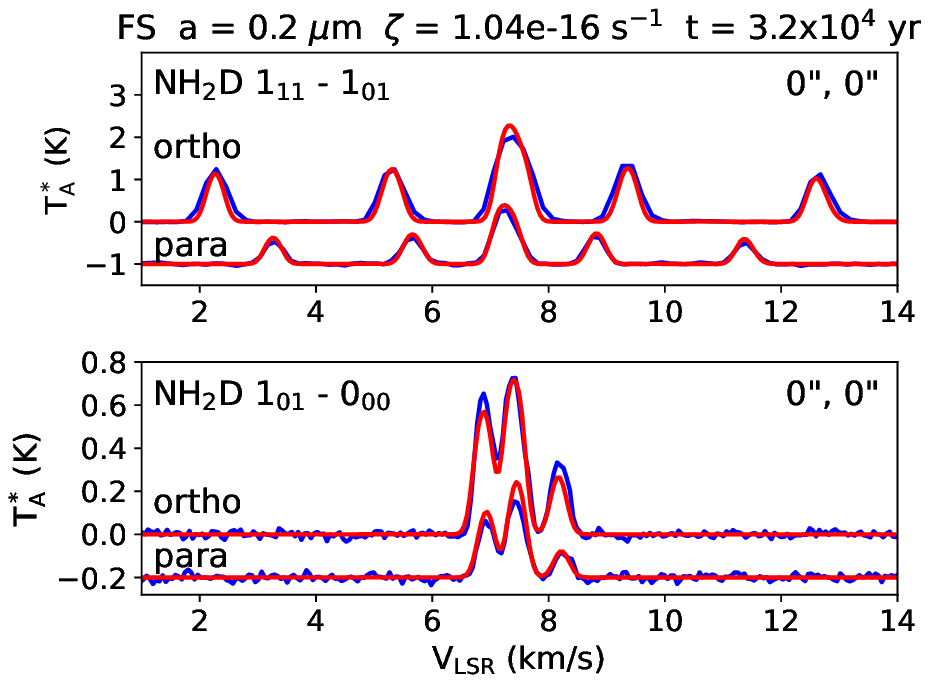}
\end{picture}}

\end{picture}  
\caption[]{Observed (blue) and simulated (red) $\dammo$ spectra towards (0,0). Left: Simulations from the PH model with two different sets of model parameters. Right: Results from the FS model with the same parameter values. The model parameters are given at the top of each panel. The upper plot in each of the four panels shows the $\exline$ lines and the lower plot shows the $\groundline$ lines of $\odammo$ and $\pdammo$. The para-lines are shifted downwards for clarity.}
\label{sample_nh2d}
\end{figure*}

{  We first attempted to reproduce the spectra observed with APEX and IRAM using the abundance
distributions predicted by the chemistry models described above.} Almost all models tested here can reasonably well reproduce the two p$\dammo$ lines and the o$\dammoline$ line at 333\,GHz at one simulation time. Examples of simulated $\dammo$ spectra towards the centre of the core are shown in Fig.~\ref{sample_nh2d}. These are the best-fit spectra from the FS and PH models with certain combinations of $a$ and $\zeta_{\htwo}$. For models assuming $a=0.2\,\mu$m and $\zeta_{\htwo}=2.6\times10^{-17}\,\pers$ (top panel), the best-fit simulation time is $3.2\times10^5$\,yr. Like for most other models, the 86\,GHz line of the o$\dammo$ line is clearly brighter than observed whenever the three other lines agree. The discrepancy depends mainly on the relatively large telescope beam ($29\arcsec$ vs $19\arcsec$-$22\arcsec$) in the 86\,GHz observations. It suggests either that the predicted $\dammo$ distribution is overly extended or that the model underpredicts the abundances near the centre of the core. In both cases, the lines simulated with smaller beams would be excessively weak with respect to the simulated 86\,GHz line. Taking a closer look at the spectra one notices that the PH model can better reproduce both 333\,GHz lines simultaneously. In the FS model, one of these lines is either too weak or too strong when the other agrees. The same is true for ortho- and para-$\ddammo$. This difference is also seen in Figs.~\ref{x_vs_time} and \ref{ops_dfracs}. For the PH model, the curves showing the fractional abundances of $\odammo$ and $\pdammo$ cross the horizontal lines indicating the values derived from observations approximately at the same time, and the predicted o/p ratios (constantly 3 and 2) are close to the observationally derived values.  For the FS model, the best-fit times are different, and the o/p-$\dammo$ and o/p-$\ddammo$ ratios are almost reversed with respect to those in the PH model. The best-fit simulation times in Fig.~\ref{sample_nh2d} do not correspond to the times when the curves shown in Fig.~\ref{x_vs_time} cross the horizontal lines because the `observed levels' shown in the latter figure were derived assuming constant abundances. Examples of simulated abundance distributions are shown below.   

\begin{figure*}
\centering
\unitlength=1mm
\begin{picture}(160,60)(0,0)
\put(-5,0){
\begin{picture}(0,0) 
\includegraphics[width=8cm,angle=0]{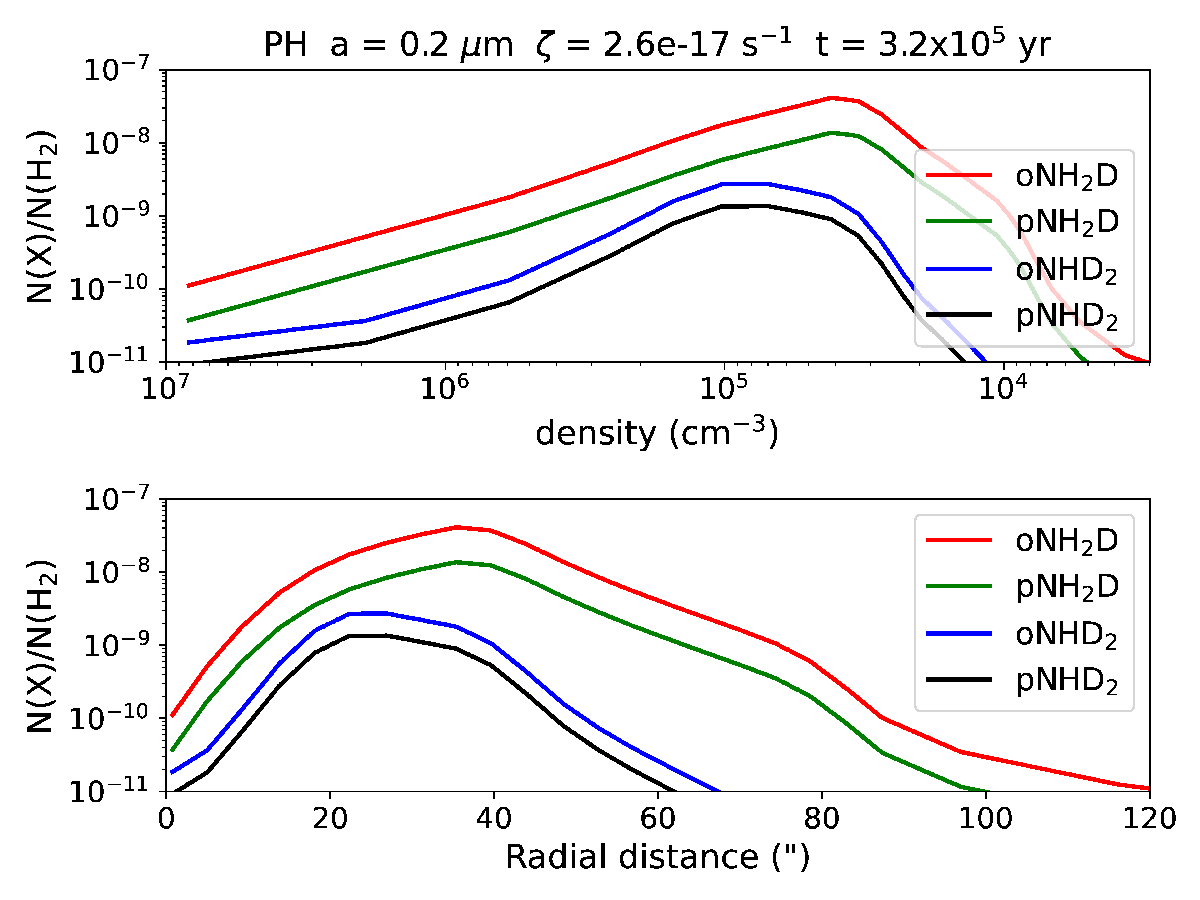}
\end{picture}}
\put(85,0){
\begin{picture}(0,0) 
\includegraphics[width=8cm,angle=0]{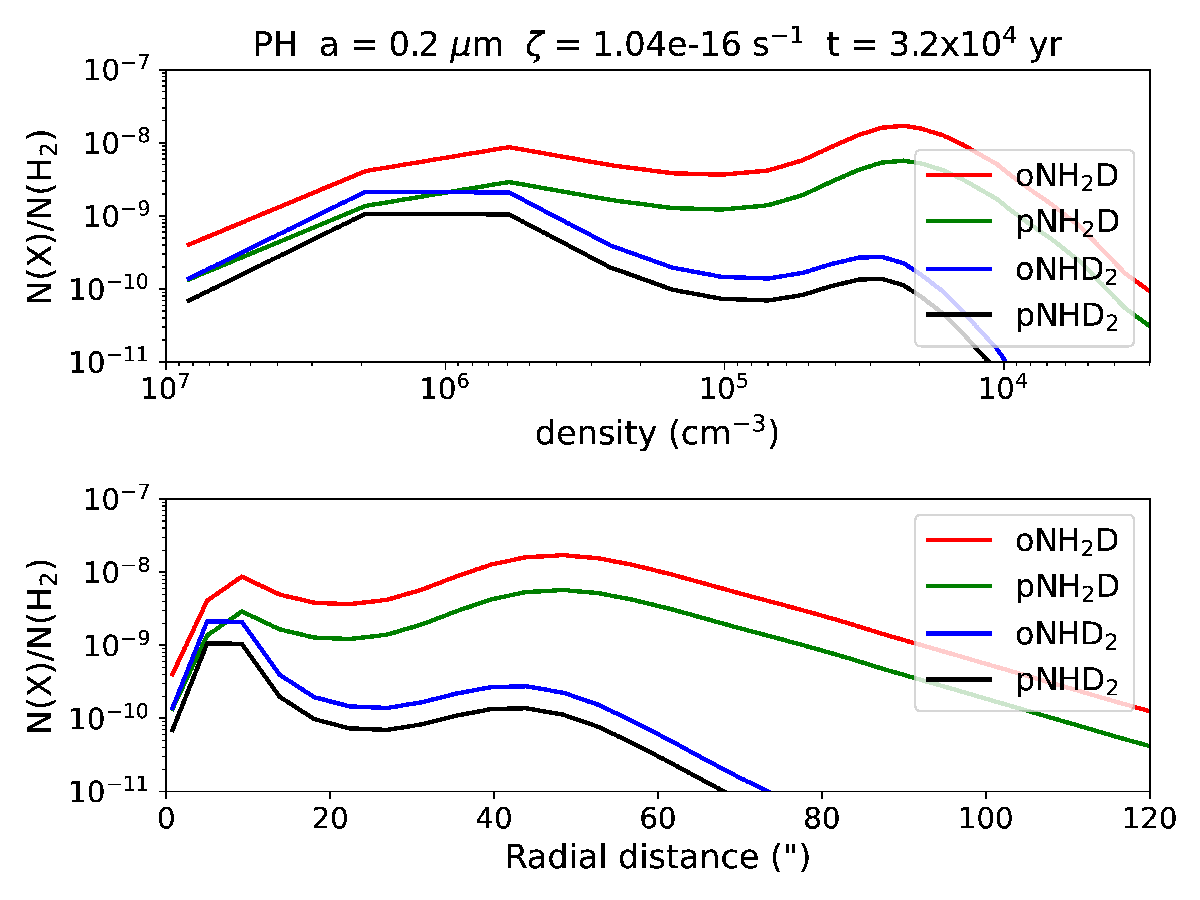}
\end{picture}}
\put(10,50){\large \bf }
\put(100,50){\large \bf }
\end{picture}  
\caption[]{Factional abundance profiles in the PH model core used to calculate the spectra shown in the left panels of Fig.~\ref{sample_nh2d}.}
\label{abundance_profiles}
\end{figure*}

The agreement between the simulated and observed 86\,GHz lines of o$\dammo$ improves with an increase in the cosmic-ray ionisation rate. In the bottom panel of Fig.~\ref{sample_nh2d}, we show the spectra obtained from the models with $\zeta_{\htwo}=1.04\times10^{-16}\,\pers$ for the grain radius $a=0.2\,\mu$m. The agreement is also good for the model with $a=0.3\,\mu$m. At high values of $\zeta_{\htwo}$ chemical reactions proceed rapidly and the 333\,GHz lines reach the observed intensities already after $3-4\times10^4$\,yr of simulation time. Also in this case, the relative intensities of the ortho- and para-lines of $\dammo$ at 333\,GHz are better reproduced by the PH model. At the best-fit time, the o/p-$\dammo$ and o/p-$\ddammo$ ratios predicted by the FS model are 2.3 and 2.9, respectively, while in the PH model they are always 3.0 and 2.0. Figure~\ref{abundance_profiles} shows the abundance profiles of $\dammo$ and $\ddammo$ in the core models used for the left-hand side of Fig.~\ref{sample_nh2d}. One can see that in the model with a high cosmic-ray ionisation rate, the abundances of $\dammo$ and $\ddammo$ are higher near the centre of the core than in the model with a lower $\zeta_{\htwo}$. In fact, the abundances of ortho- and para-$\dammo$ are almost constant in a wide range of densities and radial distances greater than $\sim 10\arcsec$ from the centre. 

\begin{figure}
\centering
\unitlength=1mm
\begin{picture}(80,105)(0,0)
\put(0,55){
\begin{picture}(0,0) 
\includegraphics[width=7cm,angle=0]{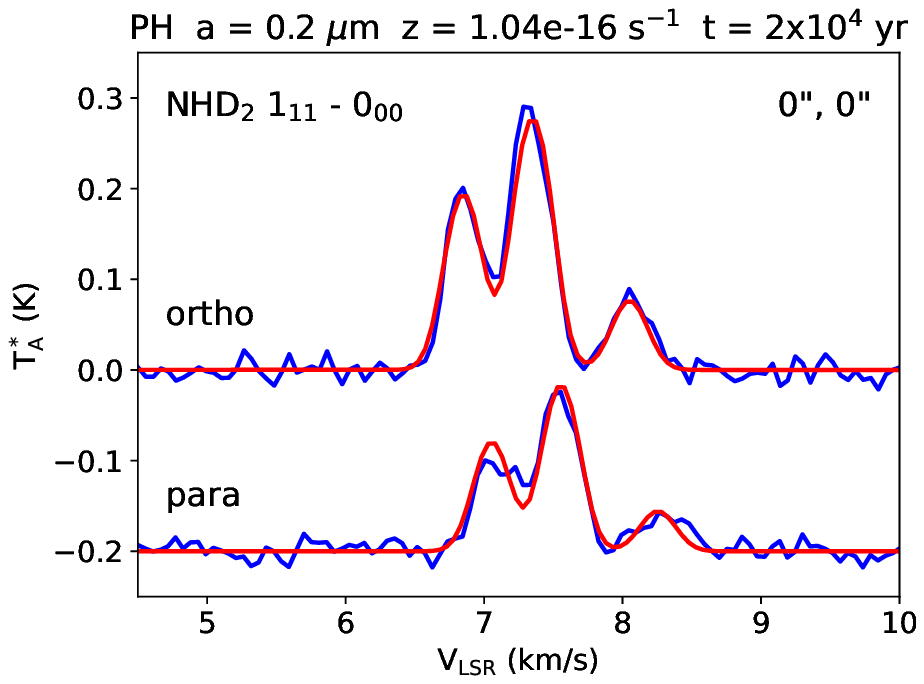}
\end{picture}}
\put(0,0){
\begin{picture}(0,0) 
\includegraphics[width=7cm,angle=0]{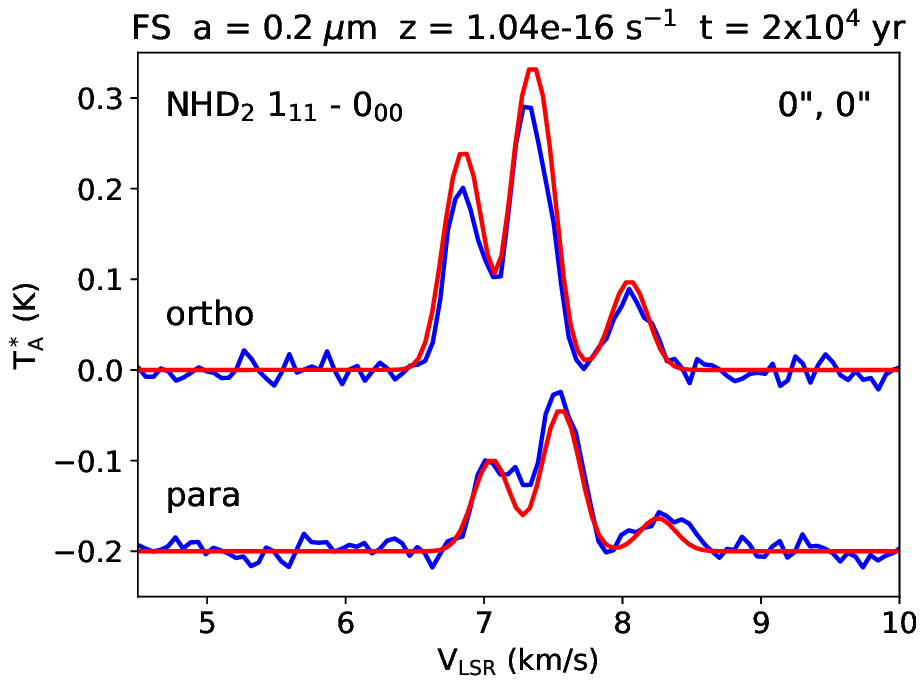}
\end{picture}}
\end{picture}  
\caption[]{Observed and simulated $\ddammo$ spectra towards the column density peak of L1544. The red lines show simulations from the PH (top panel) and FS (bottom panel) reaction schemes. The same assumptions about the effective grain radius and the cosmic-ray ionisation rate have been used ($a=0.2\,\mu$m, $\zeta_{\htwo}=1.04\times10^{-16}\,\pers$) in the two models. The simulation time has been chosen to provide an approximate agreement with the p$\ddammo$ spectrum.}
\label{sample_nhd2}
\end{figure}

The ortho- and para-$\ddammo$ lines at 335\,GHz can be fitted with models that also reproduce the 333\,GHz lines of $\dammo$. However, as suggested in Fig.~\ref{x_vs_time}, $\ddammo$ reaches the best-fit abundance earlier than $\dammo$. We think the difference is attributable to the static core model, which neglects the contraction of the core at the same time as the chemistry is evolving. This makes it difficult to reach agreement with observations simultaneously for two different species. The simulations of $\ddammo$ spectra from the models that approximately reproduce all four $\dammo$ spectra ($a=0.2\,\mu$m, $\zeta_{\htwo}=1.04\times10^{-16}\,\pers$) are shown in Fig.~\ref{sample_nhd2}. 

\section{Simulations with an ad hoc abundance profile}
\label{updown}

The agreement between the simulated and observed spectra reached by increasing the cosmic-ray ionisation rate to $\zeta_{\htwo} \sim 10^{-16}\,\pers$ is not satisfactory because a clearly lower average value of $\zeta_{\htwo}$ ($\sim 3\times10^{-17}\,\pers$) has been estimated in L1544 based on observations of molecular ions \citep{2021A&A...656A.109R}. On the other hand, the abundance profile of $\dammo$ predicted by the high-ionisation-rate model is likely to have an element of truth.  In the inner part of the core, it agrees with the ALMA observations and the modelling results of \cite{2022ApJ...929...13C}, which show that p$\dammo$ becomes heavily depleted within $10\arcsec$ (1700 au) from the centre of L1544 (see their Fig.~5). We ran another series of simulations assuming that the fractional abundances of ortho- and para-$\dammo$ and $\ddammo$ are constant in the density range $n_{\htwo}=5\times10^4-5\times10^5\,\percc$ (corresponding to radial distances $10\arcsec-30\arcsec$), but decrease both towards the centre and the outer {  parts. Here, the simulations included both the single-dish observations with APEX and IRAM and the p$\dammo$ map observed with ALMA. We first assumed that fractional abundances are proportional to $n^{-1}$ and $n$ in these two regions. A similar distribution was used to model the observations of $\dammo$ and $\ddammo$ in the pre-stellar cores H-MM1 and Oph\,D in Ophiuchus \citep{2024A&A...682A...8H}, except that it was assumed that the density threshold of the region dominated by accretion is lower. The high threshold used in L1544 is based on the observations of \cite{2022ApJ...929...13C}. Although the APEX and IRAM spectra were not very sensitive to the slope of the density dependence in the inner part of the core, to reproduce the ALMA observations it was necessary to assume that the decrease in the $\dammo$ abundance near the centre is steeper than $n^{-1}$. We obtained rather a good agreement with all the observed spectra by assuming that the fractional $\dammo$ and $\ddammo$ abundances are proportional to $n^{-2}$ in the inner parts of the core. The steepness of the slope does not affect the o/p ratios obtained from the APEX data.}    

The fractional abundances of o$\dammo$, p$\dammo$, o$\ddammo$, and p$\ddammo$ (in the flat region) were independently varied and the simulated spectra of the spheroidal core model were compared with observations.  The best-fit abundances and their uncertainties were estimated by evaluating the $\chi^2$ statistic as explained in \cite{2024A&A...682A...8H}. The results of these simulations are presented in Table~\ref{X_updown}. The simulated spectra from the best-fit model towards the seven positions observed with APEX/LAsMA are shown in Figs.~\ref{lasma_dammo} and \ref{lasma_ddammo}, together with the observed spectra at these positions. {  The observed and simulated ALMA spectra of p$\dammo$ along two orthogonal strips near the centre of the core are shown in Fig.~\ref{pnh2d_alma}. }

\begin{table}
      \caption[]{Fractional abundances, $X$, and abundance ratios in L1544 derived using a model where the abundances are constant in the density range  $n_{\htwo}=5\times10^4-5\times10^5\,\percc$ but decrease towards higher and lower densities.}

\begin{tabular}{lcc}
 & Maximum & Average \\
 & ($\times 10^{-9}$)$^a$ &  ($\times 10^{-9}$)$^a$ \\ \hline
           \noalign{\smallskip}
$X({\rm o}\dammo)$  & $9.26\pm0.06$ & $3.74\pm0.02$\\
$X({\rm p}\dammo)$  & $3.25\pm0.04$ & $1.31\pm0.02$\\
$X({\rm o}\ddammo)$ & $2.23\pm0.03$ & $0.90\pm0.01$\\
$X({\rm p}\ddammo)$ & $1.06\pm0.02$ & $0.43\pm0.01$\\ \hline
          \noalign{\smallskip}
        $X(\dammo)$ & $12.5\pm0.1$ &  $5.05\pm0.03$\\ 
        $X(\ddammo)$ & $3.30\pm0.03$ & $1.33\pm0.01$ \\ \hline \hline
        \noalign{\smallskip}
        $\dammo/\dammo$ & \multicolumn{2}{c}{$0.264\pm0.003$} \\
        \noalign{\smallskip}
        o/p$\dammo$ &  \multicolumn{2}{c}{$2.85\pm0.05$} \\
        o/p$\ddammo$ & \multicolumn{2}{c}{$2.10\pm0.06$} \\ \hline          
\end{tabular}

 $^a$ Factor of the fractional abundances, $X.$
 
\label{X_updown}
\end{table}

\section{Discussion}
\label{discussion}

\subsection{Distributions of $\dammo$ and $\ddammo$ in L1544}
\label{distribution}

The efforts to simulate the observed $\dammo$ and $\ddammo$ lines towards L1544 have shown that it is fairly easy to construct a model that approximately reproduces the observed $\dammo$ lines at 333\,GHz and the p$\dammo$ line at 110\,GHz, and the lines of $\ddammo$ at 335\,GHz. Because the beam sizes used to observe the aforementioned lines are similar ($\sim20\arcsec$) their relative intensities do not strongly depend on the spatial extent of the molecular distribution as long as the ortho and para modifications of $\dammo$ and $\ddammo$ follow the same pattern.  However, we encounter severe problems when including the 86\,GHz line of o$\dammo$, which was observed with a larger beam ($\sim29\arcsec$). For most models, the 86\,GHz line is too strong whenever the intensities of the other lines agree with the observations, indicating that the abundance of $\dammo$ is excessively high in the outer parts of the core relative to that in the central parts. Because the chemistry model predicts heavy depletion of molecules in the inner parts of the core as a result of accretion onto the grains, it is natural to suspect that the depletion is too efficient in the model. One way to decrease accretion is to increase the effective grain radius, $a$, which reduces the total surface area of the grains. Simulations with constant grain radii of $a=0.2$ and $0.3\,\mu$m gave better results than our standard assumption $a=0.1\,\mu$m. We also performed simulations with a varying grain radius according to the coagulation model of \citet[with icy grains processed for $10^5$\,years]{2009A&A...502..845O}, but these did not result in significant changes relative to our standard model. This is probably because in the varying grain size model the effective  radius  exceeds $a=0.3\,\mu$m only in a few central shells of the core model, within an angular distance of $5\arcsec$ from the centre. The 86\,GHz and 333\,GHz lines of o$\dammo$ could not be fitted simultaneously with any of the grain size distributions tested if the other model parameters were left unchanged.

By increasing the cosmic-ray ionisation rate, from the standard value $\zeta_{\htwo}=1.3\times10^{-17}$ by a factor of 8, to $\zeta_{\htwo}=1.04\times10^{-16}\,\pers$, we achieved, at an early simulation time, spatial distributions of $\dammo$ and $\ddammo$ that reproduce all of the observed lines more or less simultaneously. An increased cosmic-ray ionisation rate speeds up gas-phase chemical reactions leading to high abundances of $\dammo$ and $\ddammo$ in the inner parts of the core before depletion takes effect. The resulting  distribution of $\dammo$ is fairly flat at the beginning, except for a drop near the very centre (Fig.~\ref{abundance_profiles} right). The shape of the distribution in the inner parts of the core agrees with the ALMA observations and modelling results of \cite{2022ApJ...929...13C}, which show that p$\dammo$ becomes heavily depleted within $10\arcsec$ (1700 au) from the centre of L1544 (see their Fig.~5). As discussed in this paper, the detection of such a small, almost complete freeze-out zone requires interferometric observations.  That the present chemical model seems to overpredict molecular depletion and that single-dish observations can be modelled assuming constant fractional abundances have been previously noted by \cite{2025A&A...694A..27S}. In the present study, increasing the cosmic-ray ionisation rate is a way to achieve high abundances of $\dammo$ and $\ddammo$ in the inner parts of the core, but we are not sure how seriously the value $\zeta_{\htwo}\sim 10^{-16}\,\pers$ should be taken. It would agree with the estimate of \cite{2019ApJ...884..176I} based on the gas temperature measurements in L1544, and a model for the equilibrium gas temperature and size-dependent dust temperature. This model assumes the canonical grain size distribution of \cite{1977ApJ...217..425M} with $dn(a)/da \propto a^{-3.5}$ in the range $a_{\rm min} \leq a \leq a_{\rm max}$, where $a_{\rm min}=5$\,nm and $a_{\rm max}=250$\,nm. As noted in \cite{2021A&A...656A.109R}, this size distribution is unlikely to be valid in dense molecular clouds where the smallest grains are eliminated by coagulation \citep{2020A&A...641A..39S}, and taking this into account, the estimate from the theory of \cite{2019ApJ...884..176I} is lowered by a factor of 3. The revised value agrees with the results of \cite{2021A&A...656A.109R} giving an average $\zeta_{\htwo}$ of $3\times10^{-17}\,\pers$ in L1544. The abundances of the molecular ions observed by \cite{2021A&A...656A.109R} are highly dependent on the cosmic-ray ionisation rate, and the estimate of these authors is probably more trustworthy than the high value obtained here. The lower value would be in line with recent estimates of the cosmic-ray ionisation rate in nearby diffuse clouds ($<10^{-16}\,\pers$; \citealt{2024ApJ...973..142O}; \citealt{2024ApJ...973..143N}). These are based on updated gas density estimates from three-dimensional extinction maps and the excitation of diatomic carbon. On the other hand,  one could expect a local enhancement of the cosmic-ray ionisation rate in the Taurus molecular cloud complex, including L1544, because of their apparent location in the intersection of large-scale structures known as the Local Bubble and the Per-Tau Shell (\citealt{2022Natur.601..334Z}; \citealt{2021ApJ...919L...5B}). As discussed in \cite{2022Natur.601..334Z}, these structures are likely to have been generated by multiple supernova events in the past. Shock acceleration in the blast waves of supernova remnants is a major source of cosmic rays (e.g., \citealt{2019PhRvL.123g1101D} and references therein). 

In addition of the cosmic-ray ionisation rate, there are several other parameters in the chemistry model that influence the accretion and desorption of molecules and thereby their radial distribution in a dense core. However, the objective of the present study is to determine the o/p ratios of $\dammo$ and $\ddammo$.  In what follows, we leave the difficulties in modelling the spatial distribution of these molecules aside and concentrate on their spin ratios.

\subsection{Ortho/para ratios} 
\label{op_ratios}

Judging from the difficulty in fitting the o$\dammo$ line at 86\,GHz together with the other lines of $\dammo$, we think that observations of the ground-state lines of $\dammo$ and $\ddammo$ at 333 and 335\,GHz provide the most reliable estimates for the o/p ratios of these molecules. In addition to the fact that the ground-state lines are observed with the same beam, their relative intensities are free from calibration issues, because they lie close in frequency and are observable in the same spectrometer band. As can be seen in the spectra shown in Figs.~\ref{sample_nh2d} and \ref{sample_nhd2}, simulations from the FS reaction scheme underpredict the intensity of the o$\dammoline$ when the p$\dammoline$ line agrees, and conversely overpredict the o$\ddammoline$ intensity when p$\ddammoline$ fits. Simulations from the PH reaction scheme give reasonably good agreement with both ortho- and para-lines at the same time. The differences between the simulated line intensities from the PH and FS reaction schemes are not large, but the FS model produces systematically discrepant line ratios of the ortho- and para-species. The line simulations described in Sect.~\ref{updown}, which used a fixed fractional abundance profile but let the absolute values float freely, support the PH scenario. The best-fit fractional abundances imply the spin ratios o/p-$\dammo=2.85\pm0.05$ and o/p-$\ddammo=2.10\pm0.06$ (taking into account $1\sigma$ errors). This means that it is reasonable to assume that the o/p ratios of $\dammo$ and $\ddammo$ in L1544 are simply determined by the nuclear spin statistical weights. We arrived previously at the same conclusion in the Ophiuchus cores H-MM1 and Oph\,D.  This also agrees with the earlier results of \cite{2016MNRAS.457.1535D} towards the dense cores B1b and I16293E. 

Based on the arguments presented in \cite{2024A&A...682A...8H}, we assume that the o/p ratios of $\dammo$ and $\ddammo$ observed in the gas are not inherited from molecules released from grains, but result from ion-molecule reactions in the gas, which thus must be regarded as PH reactions rather than reactions where the H and D nuclei are completely scrambled. The viability of mixing hydrogen or deuterium atoms in reactions leading to deuterated ammonia was briefly discussed in \cite{2013JPCA..117.9800R} and \cite{2017A&A...600A..61H}, with reference to theoretical calculations of \cite{1991JChPh..94.7842H} and \cite{1992JChPh..97.1191I} of the potential energy surface of the system $[\ammo..\htwo/{\rm HD/D_2}]^+$, which is the dominant pathway to the ammonium ion ${\rm NH_4^+}$. As discussed in \cite{1992JChPh..97.1191I}, interchange of H or D atoms is possible through certain transition structures of the reaction complex, some of which have relatively low energies while others constitute substantial barriers. However, in some cases, a barrier can be avoided due to the nearly free internal rotation of the ${\rm NH_4^+}$ moiety in the exit channel complex $[{\rm NH_4..H}]^+$. We are not aware of theoretical calculations that would address the reactions $\ammo+\htwodplus/\dtwohplus/\dthreeplus$, which probably produce most of the deuterated ammonia in dense cores. We note that when only one hydrogen or deuterium atom is transferred from one molecule to another, the spin ratios maintain their statistical values even if nuclei are mixed within the two parts of the entrance and exit channel complexes. 

As discussed in \cite{2019A&A...631A..63S}, the choice of the reaction scheme also affects the deuterium fractionation in addition to the spin ratios. Reactions where two deuterium atoms are transferred, such as $\ammo+\dtwohplus\rightarrow \ddammo+\htwo$, are possible in the FS scheme but not allowed in the PH scheme (see Fig.~1 in \citealt{2019A&A...631A..63S}). In the present simulations, the PH model predicts higher $\dammo/\ammo$ ratios and lower $\ddammo/\dammo$ and $\ndthree/\ddammo$ ratios than the FS model, especially at late times ($t\goa10^6$\,yr; see Fig.~\ref{ops_dfracs}). Because the abundance of $\ddammo$ peaks earlier than that of $\dammo$ (see Fig.~\ref{x_vs_time}), the predicted $\ddammo/\dammo$ ratio generally does not correspond to the observed value at the time when the $\dammo$ spectra agree with the observations. In contrast to previous simulations of \cite{2019A&A...631A..63S}, the ratio $\dammo/\ammo$ remains low ($\sim 0.1$) also in the PH model. The present chemistry models include chemical desorption, which enhances the replenishment of the gas with molecules formed in the icy mantles of grains, where deuterium fractionation is weaker. These simulations also show that the molecules released from the grains are quickly reprocessed in the gas, and the spin ratios are determined by the ion-molecule reactions no matter how efficient the desorption is \citep{2024A&A...682A...8H}. As evident in Fig.~\ref{ops_dfracs}, the predicted deuterium fractionation ratios of ammonia vary significantly during the probable lifetime of a dense core ($\loa10^6$\,yr). The figure suggests that fractionation ratios can be useful for chemical dating, while spin ratios do not serve this purpose.   

\section{Conclusions}

We have determined the o/p ratios of the $\dammo$ and $\ddammo$ molecules in one of the most sheltered and coldest places in the local Universe, the inner parts of the pre-stellar core L1544 in the Taurus molecular cloud complex. In accordance with similar measurements towards two starless cores in Ophiuchus, the o/p ratios of both molecules are found to correspond to the nuclear spin statistical weights, o:p-$\dammo=3:1$ and o:p-$\dammo=2:1$. Because the ratios are likely to be determined by gas-phase ion-molecule reactions, these reactions are best described as single-proton or single-deuteron hops that maintain the statistical spin ratios. Ammonia formation on grain surfaces is predicted to proceed through H or D atom additions that also produce ortho- and para-species according to their statistical weights, so it seems reasonable to assume that the spin ratios of ammonia and its deuterated isotopologues are always statistical. Another consequence of adopting the PH reaction scheme is that singly deuterated ammonia is favoured at the cost of $\ddammo$ and $\ndthree$. This scheme therefore leads to fractionation ratios $\dammo/\ammo$, $\ddammo/\dammo$, and $\ndthree/\ddammo$ that are different from those predicted by the FS scheme.    

\begin{acknowledgements}
The authors thank APEX staff for performing the observations, the anonymous reviewer for helpful comments on the manuscript, and the Max Planck Society for financial support. 
\end{acknowledgements}

\bibliographystyle{aa} 

\bibliography{bibliography.bib} 

\begin{appendix}

  \onecolumn
  
\section{Observed and simulated APEX and IRAM spectra}

\begin{figure*}[h!]
\centering
\unitlength=1.4mm
\begin{picture}(160,110)(0,0)
\put(40,40){
\begin{picture}(0,0) 
\includegraphics[width=5.5cm,angle=0]{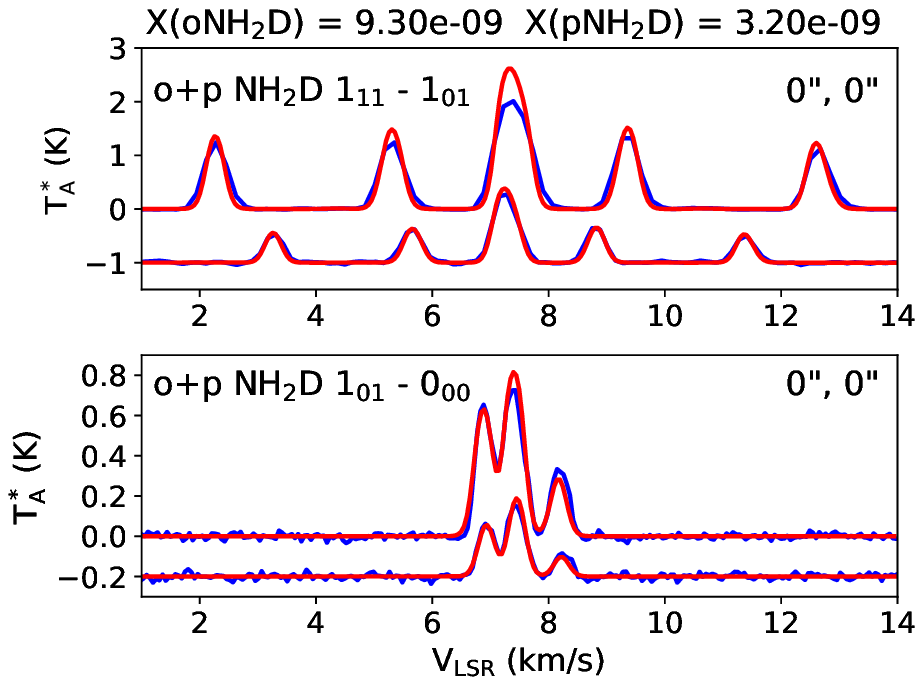}
\end{picture}}
\put(19,4){
\begin{picture}(0,0) 
\includegraphics[width=5.5cm,angle=0]{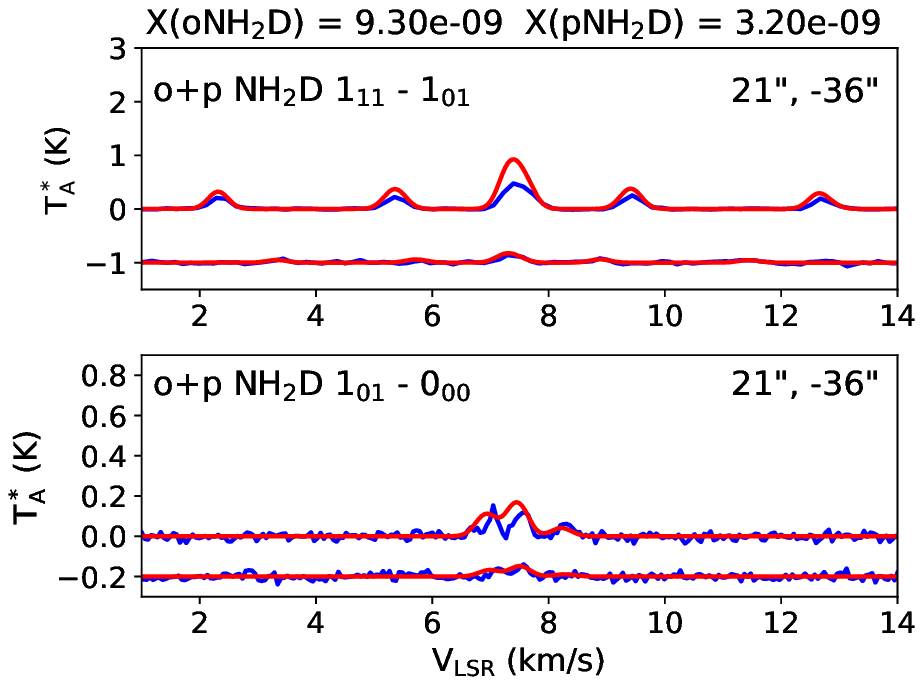}
\end{picture}}
\put(-1,41){
\begin{picture}(0,0) 
\includegraphics[width=5.5cm,angle=0]{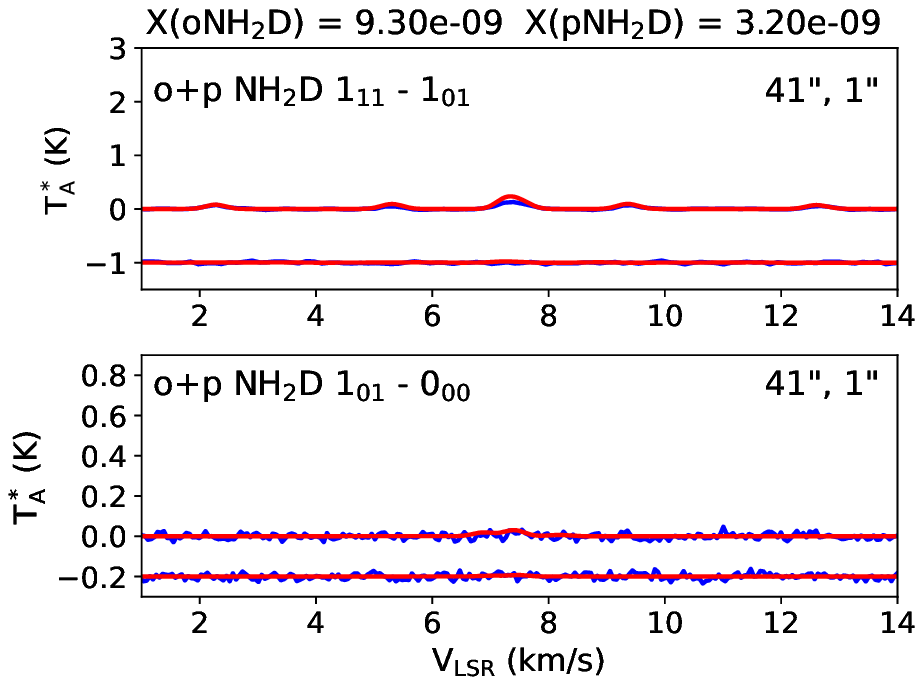}
\end{picture}}
\put(20,75){
\begin{picture}(0,0) 
\includegraphics[width=5.5cm,angle=0]{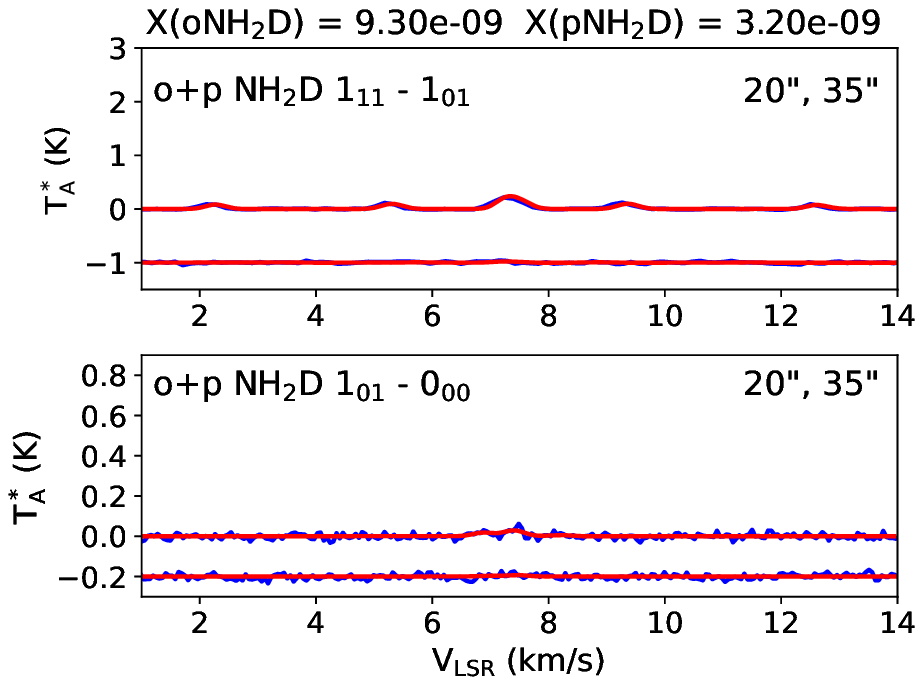}
\end{picture}}
\put(63,74){
\begin{picture}(0,0) 
\includegraphics[width=5.5cm,angle=0]{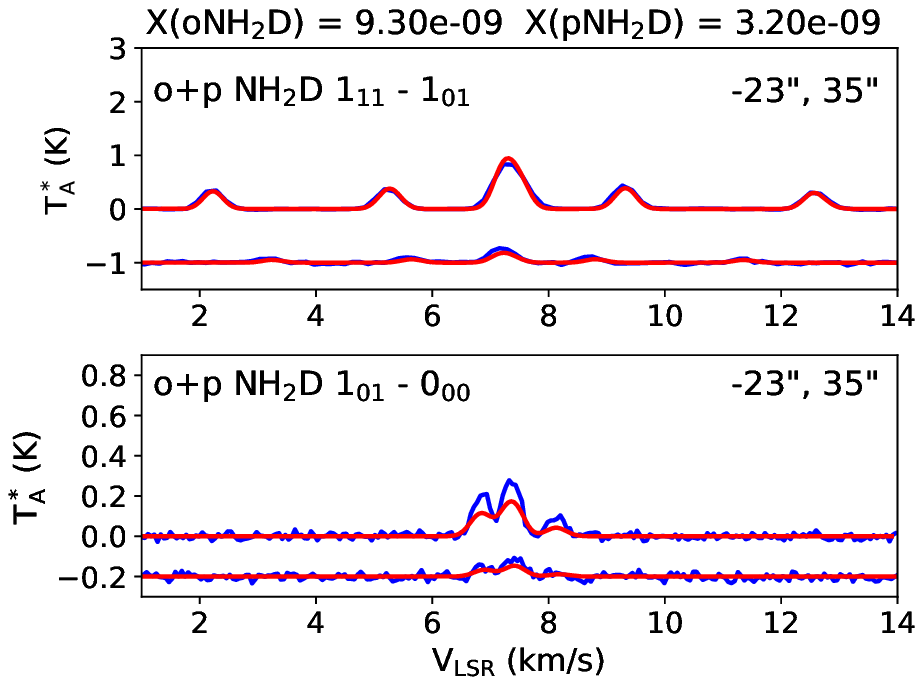}
\end{picture}}
\put(80,38){
\begin{picture}(0,0) 
\includegraphics[width=5.5cm,angle=0]{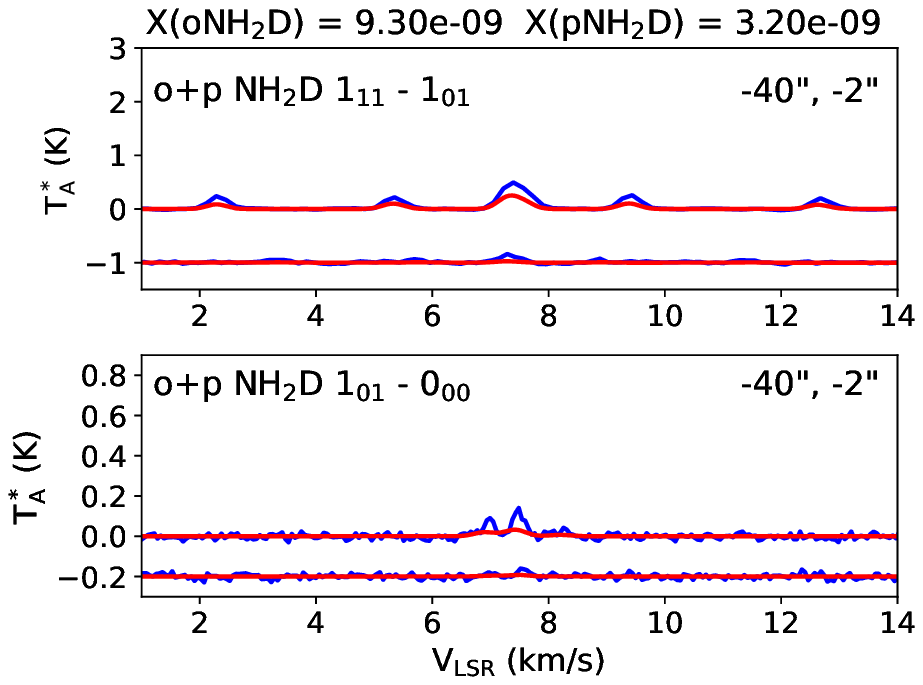}
\end{picture}}
\put(59,3){
\begin{picture}(0,0) 
\includegraphics[width=5.5cm,angle=0]{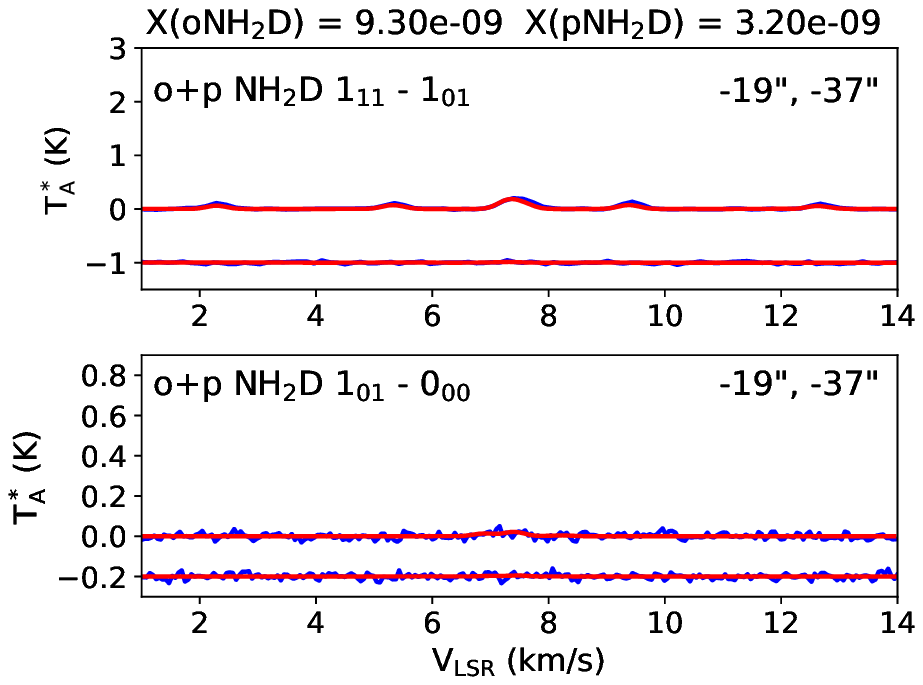}
\end{picture}}
\end{picture}
\caption{$\dammo$ spectra observed with the APEX and IRAM 30\,m telescopes (blue lines). The positions correspond to the seven pixels of the LAsMA array. Upper panels: $\exline$ spectra from IRAM 30\,m. Lower panels: $\groundline$ from APEX. Simulated spectra from the model discussed in Sect.~\ref{updown} are shown in red.}
\label{lasma_dammo}
\end{figure*}

\begin{figure*}[h!]
\centering
\unitlength=1.4mm
\begin{picture}(160,110)(0,0)
\put(40,40){
\begin{picture}(0,0) 
\includegraphics[width=5.5cm,angle=0]{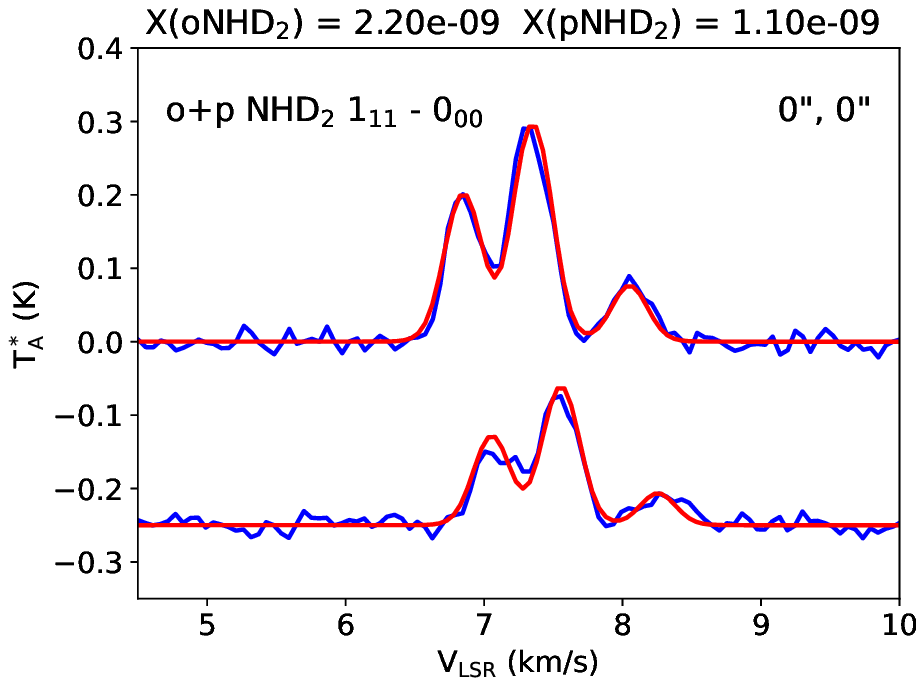}
\end{picture}}
\put(19,4){
\begin{picture}(0,0) 
\includegraphics[width=5.5cm,angle=0]{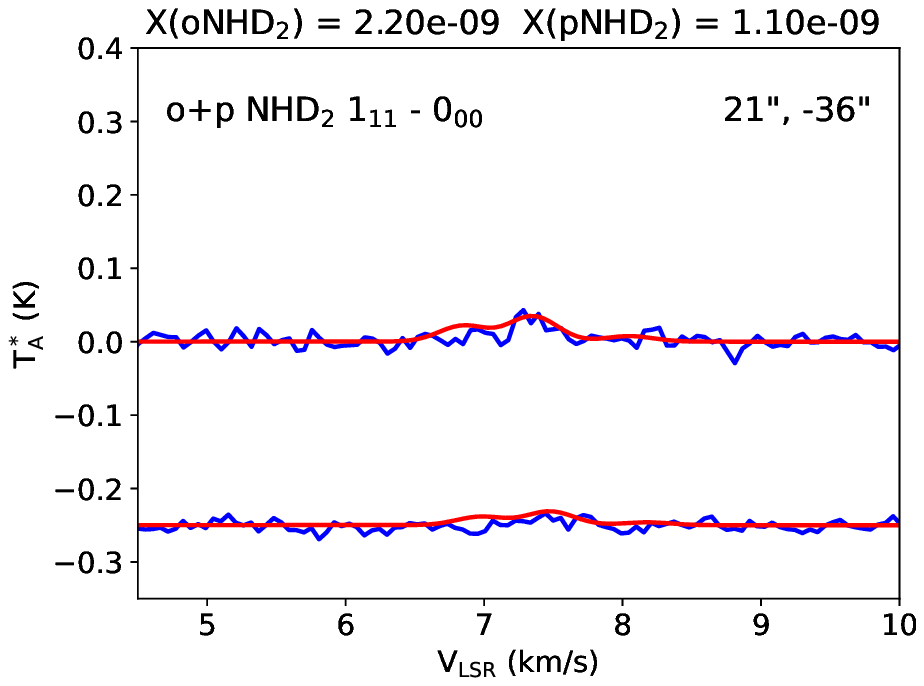}
\end{picture}}
\put(-1,41){
\begin{picture}(0,0) 
\includegraphics[width=5.5cm,angle=0]{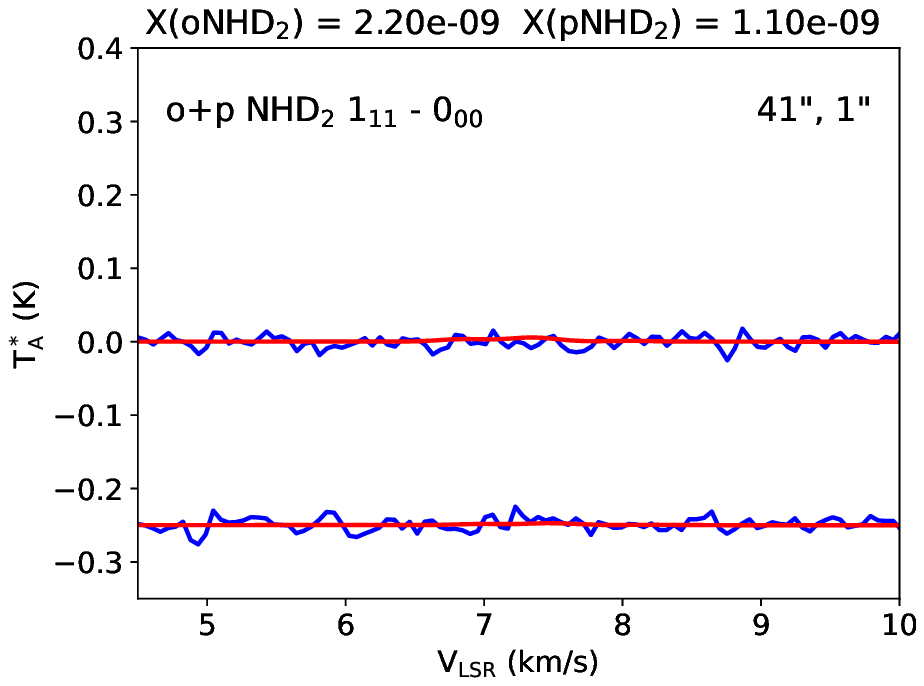}
\end{picture}}
\put(20,75){
\begin{picture}(0,0) 
\includegraphics[width=5.5cm,angle=0]{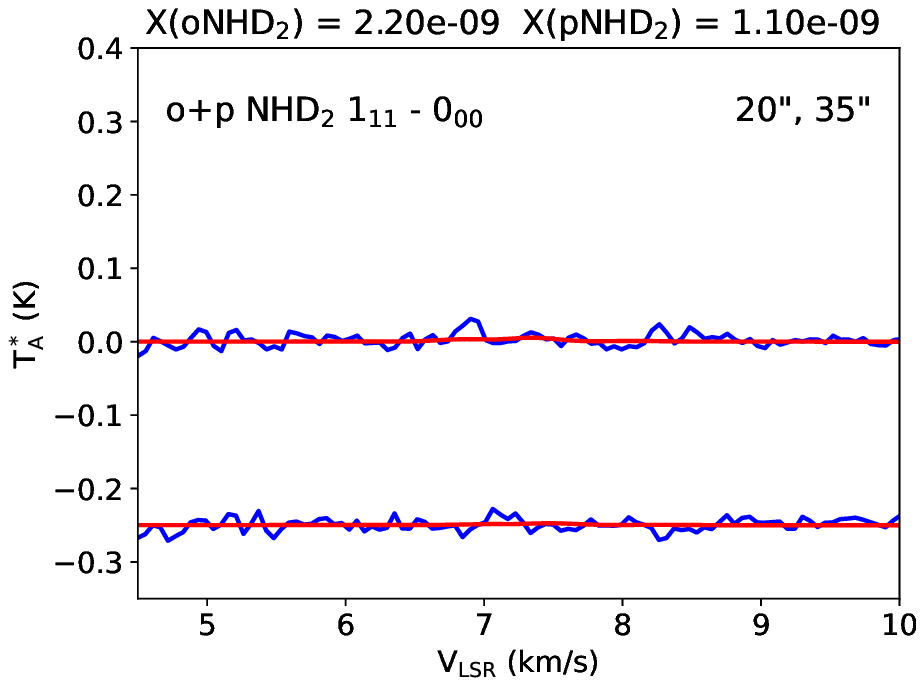}
\end{picture}}
\put(63,74){
\begin{picture}(0,0) 
\includegraphics[width=5.5cm,angle=0]{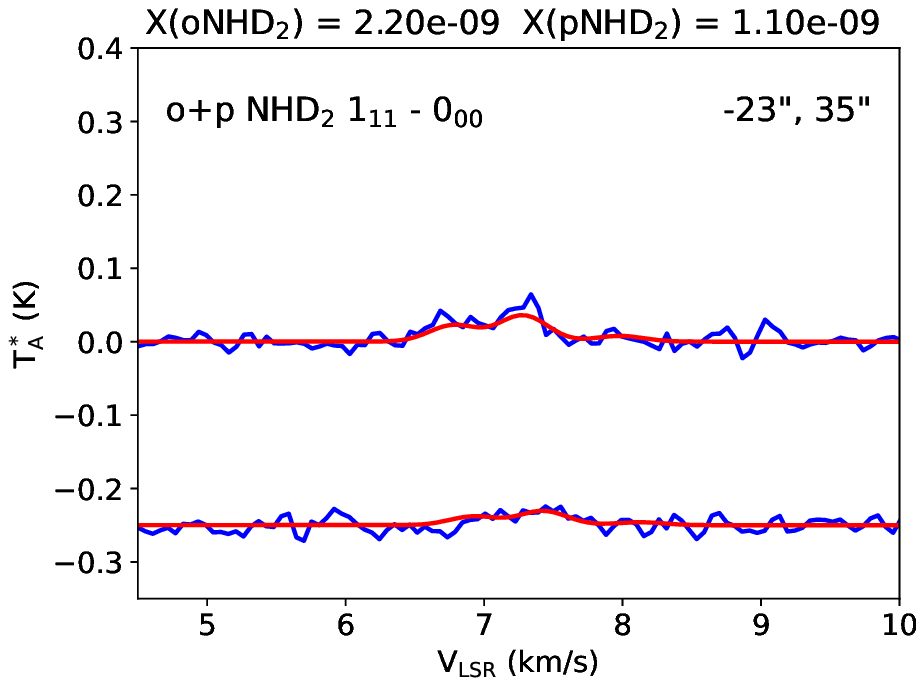}
\end{picture}}
\put(80,38){
\begin{picture}(0,0) 
\includegraphics[width=5.5cm,angle=0]{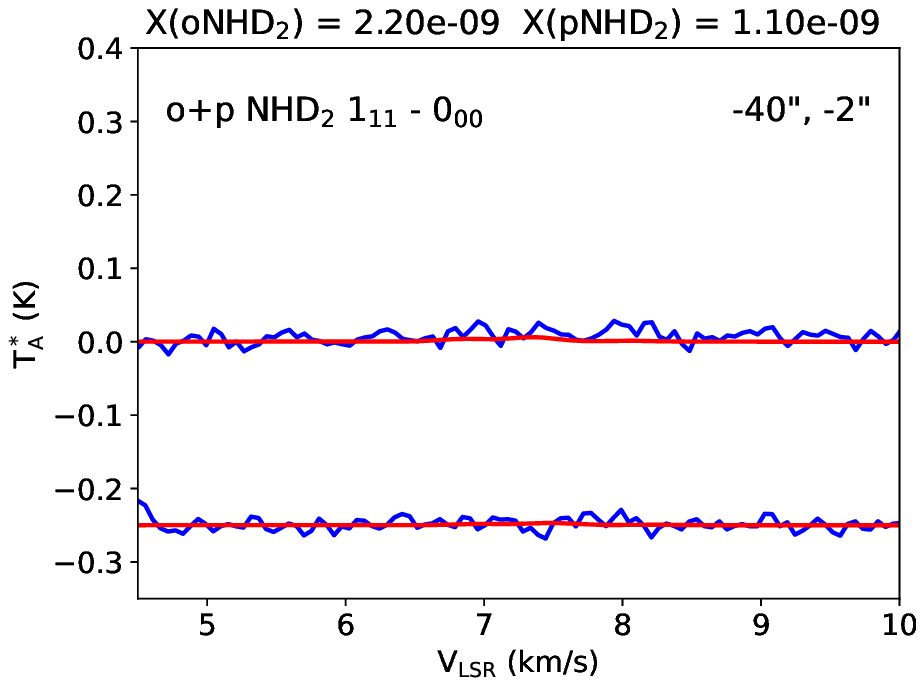}
\end{picture}}
\put(59,3){
\begin{picture}(0,0) 
\includegraphics[width=5.5cm,angle=0]{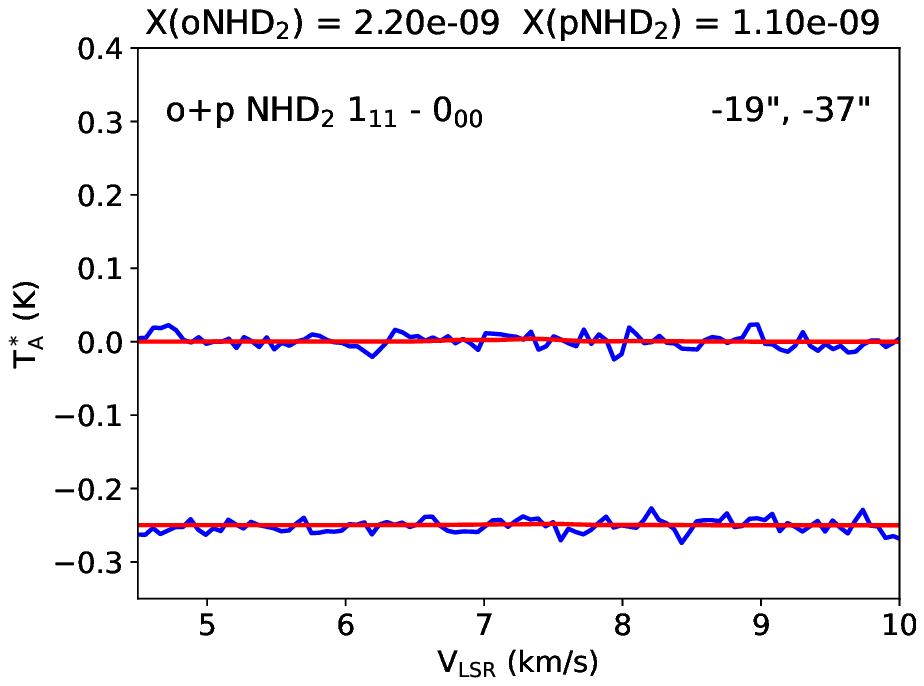}
\end{picture}}
\end{picture}
\caption{ $\exgroundline$ spectra of $\oddammo$ and $\pddammo$ observed with the LAsMA array (blue lines). The simulations from the model discussed in Sect.~\ref{updown} are shown in red.}
\label{lasma_ddammo}
\end{figure*}

The $\dammo$ spectra observed towards positions corresponding to the seven pixels of the LAsMA array are shown in Fig.~\ref{lasma_dammo}. These positions were observed in the $\groundline$ lines with APEX. The $\exline$ spectra are extracted from the maps observed with the IRAM 30\,m telescope. The $\exgroundline$ spectra of $\ddammo$ observed with APEX are shown in Fig.~\ref{lasma_ddammo}. In both figures, the predictions using the abundance profile described in Sect.~\ref{updown} are shown in red. In this profile, the maximum fractional abundance is assumed to be found at intermediate densities, with decreasing tendencies both towards the centre and towards the outer boundary of the core. The maximum values are given at the top of the panels. The offset from the core centre is indicated on the top right of each panel.  

\newpage

\section{Observed and simulated ALMA spectra}

\begin{figure*}[h!]
\unitlength=1mm
\centering
\begin{picture}(160,50)
\put(-17,2){
\begin{picture}(0,0) 
\includegraphics[width=6cm,angle=0]{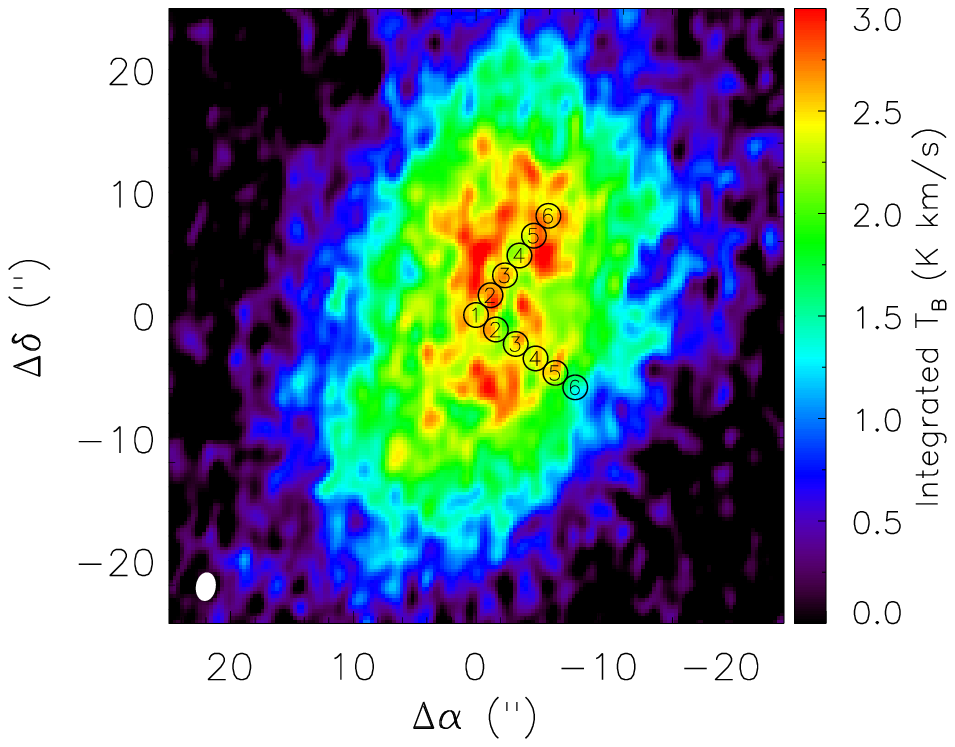}
\end{picture}}
\put(42,0){
\begin{picture}(0,0) 
\includegraphics[width=6.5cm,angle=0]{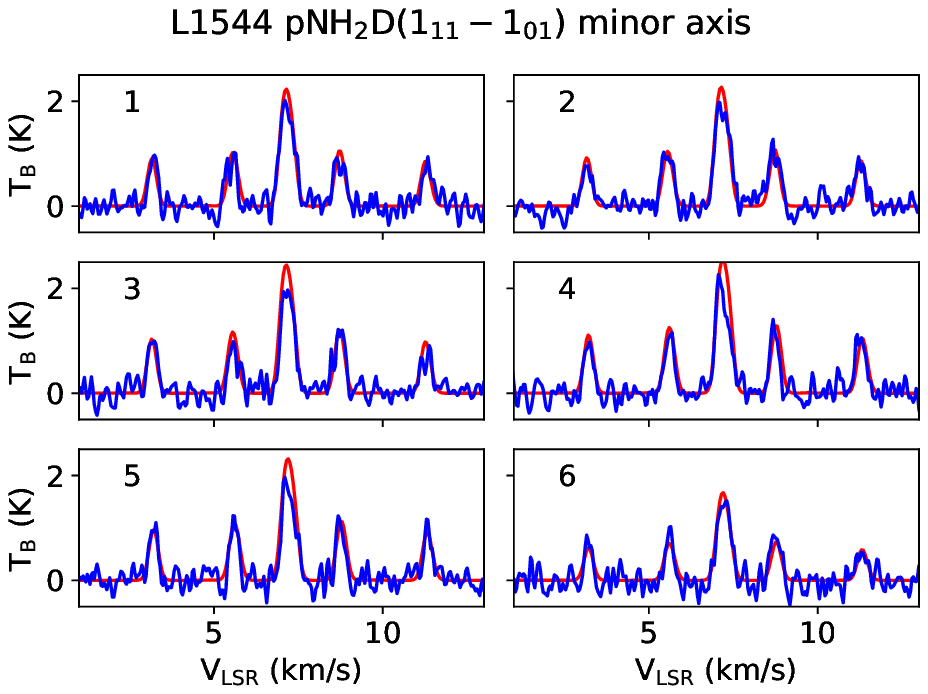}
\end{picture}}
\put(106,0){
\begin{picture}(0,0) 
\includegraphics[width=6.5cm,angle=0]{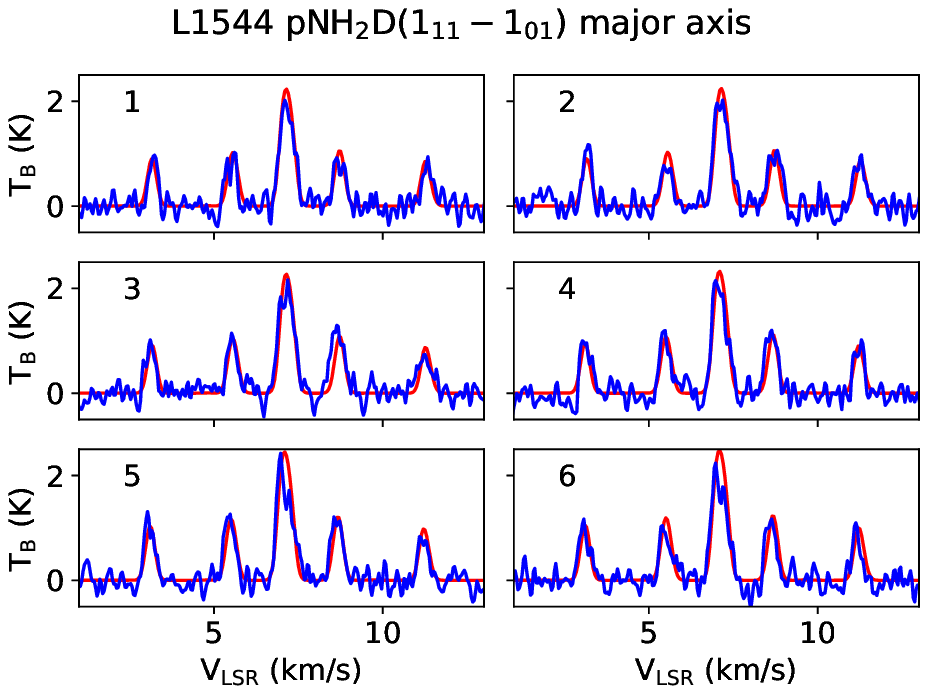}
\end{picture}}
\put(-2,46){\bf (a)}
\put(48,46){\bf (b)}
\put(113,46){\bf (c)}
\end{picture}
\label{pnh2d_alma}
\caption{   $1_{11}-1_{01}$ spectra of $\pdammo$ observed with ALMA along two orthogonal strips across L1544. The data are from \cite{2022ApJ...929...13C}. (a): Integrated line intensity map. The positions where the spectra are extracted are indicated with numbered circles. The synthesised beam is shown in the bottom left. (b): Six spectra along the minor axis of the core (from the centre to the south-west). The number in the top left of each panel refers to the circled numbers on the map. The positions are separated by $2\arcsec$.  (c): Same as panel (b) but along the major axis of the core (from the centre to the north-west). The spectra predicted from the model described in Sect.~\ref{updown} are shown in red. Unlike the integrated intensity map of panel (a), the observed spectra in panels (b) and (c) have been corrected for the primary beam response.}
\end{figure*}

{  The integrated p$\dammo(1_{11}-1_{01})$ line intensity map of L1544 and a selection of spectra observed with ALMA by \cite{2022ApJ...929...13C} are shown in Fig.~\ref{pnh2d_alma}. The two line segments where the spectra are extracted start from the column density peak (our (0,0)), and are aligned with the minor and major axis of the elliptical core as determined in Sect.~\ref{core_model}. The spectra calculated from the ad hoc model discussed in Sect.~\ref{updown} are shown in red. The position angle of the core as delineated by $\dammo$ emission ($\sim -25\degr$) differs from the angle determined by $8\,\mu$ absorption ($\sim -36\degr$). Similar figures were shown in \citet[their Figs. 3 and A.1]{2022ApJ...929...13C}, except that the strip along the main axis was directed south-east. There, the observed line emission weakens more quickly towards the core edge than in the north-west, where the spheroidal model agrees well with the observations.}    

\end{appendix}
\end{document}